\documentclass{article}

\usepackage{arxiv}

\usepackage[utf8]{inputenc} 
\usepackage[T1]{fontenc}    
\usepackage{hyperref}       
\usepackage{url}            
\usepackage{booktabs}       
\usepackage{amsfonts}       
\usepackage{nicefrac}       
\usepackage{microtype}      
\usepackage{lipsum}		
\usepackage{graphicx}
\usepackage[numbers]{natbib}
\usepackage{doi}
\usepackage{mathtools}
\usepackage{amsmath}
\usepackage{xcolor}
\usepackage{tikz}

\title{Sparse Reconstruction of Multi-Dimensional Kinetic Distributions}
\author{
{\hspace{1mm}Georgii Oblapenko$^1$},
 {\hspace{1mm}Manuel Torrilhon$^1$}, {\hspace{1mm}Michael Herty$^2$},\\
	$^1$Chair of Applied and Computational Mathematics, RWTH Aachen,
	Schinkelstrasse 52, 52062 Aachen, Germany \\
	$^2$Institute of Geometry and Applied Mechanics, RWTH Aachen,
	Templergraben 55, 52062 Aachen, Germany \\
	$^\ast$Corresponding author. E-mail: \texttt{oblapenko@acom.rwth-aachen.de}
}

\date{}

\renewcommand{\headeright}{}
\renewcommand{\undertitle}{}
\renewcommand{\shorttitle}{}

\hypersetup{
pdftitle={Sparse Reconstruction of Multi-Dimensional Kinetic Distributions},
pdfsubject={},
pdfauthor={Georgii Oblapenko, Manuel Torrilhon, Michael Herty},
pdfkeywords={Boltzmann equation, moment method, maximum entropy, sparse reconstruction, numerical optimization},
}

\begin{document}
\maketitle

\begin{abstract}
	In the present work, we propose a novel method for reconstruction of multi-dimensional kinetic distributions, based on their representation as a mixture of Dirac delta functions. The representation is found as a solution of an optimization problem. Different target functionals are considered, with a focus on sparsity-promoting regularization terms. The proposed algorithm guarantees non-negativity of the distribution by construction, and avoids an exponential dependence of the computational cost on the dimensionality of the problem.
    Numerical comparisons with other classical methods for reconstruction of kinetic distributions are provided for model problems, and the role of the different parameters governing the optimization problem is studied.
\end{abstract}

\keywords{Boltzmann equation, moment method, maximum entropy, sparse reconstruction, numerical optimization}

\section{Introduction}
\let\thefootnote\relax
\footnotetext{MSC Subject codes: 82D05, 35L60, 35Q70, 82C40}
The moment closure problem is a well-known question of estimating a moment of an unknown probability distribution function $f$ given only the
values of the lower-order moments~\cite{hamburger1944hermitian,Shohat1945problem,Aheizer1962,schmuedgen2017moment}.
Often arising from physical models characterized by a transport process of the underlying probability distribution function, the moment closure problem plays a role in multiple scientific contexts, such as rarefied gas dynamics~\cite{grad2kinetic,levermore1996moment,torrilhon2016modeling}, radiation transport~\cite{struchtrup1998number,modest2021radiative}, traffic flow modeling~\cite{marques2013kinetic,herty2020bgk}, biology~\cite{singh2006moment,gillespie2009moment}, and ocean and atmospheric sciences~\cite{milbrandt2005multimoment,yuan2012extended,koellermeier2020analysis}, to name a few.

A large number of approaches to solution of the moment problem associated with transport equations have been developed in different contexts, such as Grad's method~\cite{grad2kinetic} and its regularized version~\cite{struchtrup2003regularization}, quadrature-based approaches~\cite{mcgraw1997description,fox2008quadrature,desjardins2008quadrature,chalons2010beyond,fox2018conditional,van2021higher,huang2020stability,FoxLaurent,fox2023generalized,yilmaz2024nonlinear}, projection approaches~\cite{koellermeier2014framework}, and maximum entropy methods~\cite{levermore1997entropy,mcdonald2013affordable,abdelmalik2016moment,alldredge2019regularized,bohmer2020entropic}. Whilst many of these methods have been shown to have attractive theoretical properties (such as strong hyperbolicity of the resulting equations, entropy dissipation, numerical stability), they are mostly well-developed and studied only for the case of a one-dimensional distribution function.

In the multi-dimensional case, the question of finding a moment closure in the context of kinetic equations has been less studied,
despite the fact that multi-dimensional kinetic equations are oftentimes required for the description of various phenomena~\cite{Eftimie2018}. In the multi-dimensional case, most of the works utilize either 
the Grad closure, which is known to produce unphysical negative values of the distribution function, or the maximum-entropy approach or some version thereof~\cite{mcdonald2013affordable,bohmer2020entropic}, which guarantees the non-negativity of the reconstructed distribution function.
However, for some of these multi-dimensional formulations~\cite{bohmer2020entropic}, unlike the one-dimensional case, no convergence guarantees exist. In addition, the multi-dimensional entropy-based approaches require use of multi-dimensional quadratures, meaning their 
computational cost scales exponentially with the dimension of the domain of $f$, as tensor product grids are constructed.
A multi-dimensional quadrature-based approach was presented in~\cite{yuan2011conditional}, which however is algorithmically and computationally expensive.

In the present work, we are interested in reconstructing a non-negative distribution function from a given set of multi-dimensional moments thereof. To this end, we develop an optimization-based approach that enforces a sparse representation of $f$ in terms of a set of basis functions. By fixing the expansion weights, we automatically guarantee non-negativity of the solution and satisfaction of the zero-order moment constraint. This is contrast to other approaches, where either both the weights and velocities are not fixed~\cite{wheeler1974modified,fox2008quadrature}, or the velocities are fixed~\cite{bohmer2020entropic}. Via the use of the sparsity constraint, we can use a large pre-determined number of basis functions and yet obtain a solution that contains fewer degrees of freedom than the initial solution. This makes it possible to efficiently store the solutions and use them as starting solutions for subsequent solutions of the moment problem in time-dependent transport equations. The dimension of the solution also scales linearly with the dimension of the domain of $f$. The closure can also be used for computing both odd- and even-order moments, whereas some other closure methods are not defined in the even-order case (predicting the next moment from a set of moments of total even order)~\cite{levermore1997entropy} or lose some of their properties~\cite{cai2021moment,yilmaz2024nonlinear}. As such, the developed method can be used as a closure for multi-dimensional moment equations of arbitrary degree arising from an underlying kinetic equation. 
Additionally, such a representation provides an interesting link to particle-based approaches, and can be used for development of hybrid multi-scale particle/moment numerical schemes~\cite{crestetto2012kinetic}, as well as moment-preserving particle merging approaches~\cite{gonoskov2022agnostic,oblapenko2024non}.

We apply the approach to several model two-dimensional distributions and compare it to other existing closure methods. As such, in the present work we only focus on reconstructing given distributions and do not model the full kinetic transport problem.


\section{Multi-dimensional moment closure problem}
We define moments of a multi-dimensional distribution function $f(\mathbf{v}):\mathbb{R}^d \to \mathbb{R}_{\geq 0}$ as
\begin{equation}
    M_{k_1,\ldots,k_d} = \int  f(\mathbf{v})\prod_{l=1}^{d} v_l^{k_l} \mathrm{d} \mathbf{v} \label{eq:constraints},
\end{equation}
where $k_l \geq 0$, $l=1,\ldots,d$.

The moment problem can be defined as follows: given only the set of values $\{M_{k_1,\ldots,k_d}\}_{\sum k_l \leq N}$ (i.e. all moments of total order less or equal than $N$), we want to compute the higher-order moments $\{M_{k_1,\ldots,k_d}\}_{\sum k_l = N+1}$.

This is motivated by the multi-dimensional kinetic equation~\cite{cercignani1988}:
\begin{equation}
    \frac{\partial f}{\partial t} + \mathbf{v} \cdot \nabla f = \mathcal{Q}(f,f),\label{eq:kinetic}
\end{equation}
where $\mathbf{v} \in \mathbb{R}^d$, $\mathcal{Q}(f,f)$ is the so-called ``collision operator'', and $f$ is the so-called ``velocity distribution function'' (VDF), describing the number density of particles having a certain velocity $\mathbf{v}$. In the present work we are interested in the moment problem only, and do not consider the full kinetic equation in detail, thus, for simplicity we can assume that $\mathcal{Q} \equiv 0$. By multiplying~(\ref{eq:kinetic}) with $\prod_{l=1}^{d} v_l^{k_l}$, $k_l \geq 0$ and integrating over $\mathbf{v}$, one obtains the following equations for the moments of the distribution function $f$:
\begin{equation}
    \frac{\partial M_{k_1,\ldots,k_d}}{\partial t} + \nabla \cdot \mathcal{F}_{k_1,\ldots,k_d} = 0,\:k_l \geq 0,\: l=1,\ldots,d,\label{eq:transport}
\end{equation}
where the flux is given by
\begin{equation}
    \mathcal{F}_{k_1,\ldots,k_d} = \left(M_{k_1+1,k_2,\ldots,k_{d-1},k_d},M_{k_1,k_2+1,\ldots,k_{d-1},k_d},\ldots,M_{k_1,k_2+1,\ldots,k_{d-1}+1,k_d},M_{k_1,k_2,\ldots,k_{d-1},k_d+1}\right).
\end{equation}
As each equation for the time evolution of $M_{k_1,\ldots,k_d}$ involves higher-order moments of $f$, a closure is required, i.e. a way to compute $M_{k_1+1,k_2,\ldots,k_{d-1},k_d}$, $M_{k_1,k_2+1,\ldots,k_{d-1},k_d}$, etc~\cite{struchtrup2005}.

One of the possible approaches for solving the moment problem is by reconstructing a function $\tilde{f}$ that satisfies the moment constraints given by Eqn.~(\ref{eq:constraints}) for $k_l: \sum k_l \leq N$, and computing the moments of interest $\{M_{k_1,\ldots,k_d}\}_{\sum k_l = N+1}$ by numerical integration. However, this problem is not well-posed, as more than one $\tilde{f}$ satisfying the constraints may exist; the question of the non-negativity of the computed $\tilde{f}$ is also crucial, as negative values may lead to numerical instabilities and/or unphysical results.

\section{Numerical approach}
As throughout the paper we will be primarily dealing with the ``reconstructed''  $\tilde{f}$, we will denote is simply as $f$.

\subsection{Delta function representation}
We represent the $f(\mathbf{v}) \geq 0$, $\mathbf{v} \in \mathbb{R}^d$ as a sum of Dirac delta functions:
\begin{equation}
    f(\mathbf{v})=f(v_1,\ldots,v_d) = \sum_{i=1}^{N_p} f_i \prod_{l=1}^{d} \delta(v_l - v_{l,i}).\label{eq:f-expansion}
\end{equation}
Here $N_p$ is the number of terms in the expansion, $f_i$ are the expansion coefficients, $v_{x,i}$ and $v_{y,i}$ are the locations of the Dirac delta functions. In order to guarantee that $f(v_1, v_2) \geq 0$, the expansion coefficients $f_i$ need to be non-negative.  We note that in the representation~(\ref{eq:f-expansion}), multiple Diracs are allowed to be located at the same position.
Using the ansatz~(\ref{eq:f-expansion}), we can write the expression for moment $M_{k,l}$ as
\begin{equation}
    M_{k_1,\ldots,k_d} = \sum_i f_i \prod_{l=1}^{d} v_{l,i}^{k_l}.\label{eq:f-dirac-moments}
\end{equation}
In the one-dimensional case $d=1$, the weights $f_i$ and positions $v_{l,i}$ appearing in expression~(\ref{eq:f-expansion}) can be computed from the moments of $f$ by existing algorithms~\cite{wheeler1974modified,fox2018conditional}.
However, in the multi-dimensional case, the existence of a unique solution of the form~(\ref{eq:f-expansion}) satisfying a set of moment constraints~(\ref{eq:f-dirac-moments}) is not clear, although some approaches have been shown to be quite robust~\cite{bohmer2020entropic}.

If one fixes the positions $v_{l,i}$ and solves only for the coefficients $f_i$, the moment constraints become a system of linear equations for $f_i$:
\begin{equation}
    \begin{pmatrix}
    \prod_{l=1}^{d} v_{l,1}^{k^{(1)}_l} & \ldots & \prod_{l=1}^{d} v_{l,N_p}^{k^{(1)}_l} \\
    & \ddots & \\
    \prod_{l=1}^{d} v_{l,1}^{k^{(N_m)}_l} & \ldots & \prod_{l=1}^{d} v_{l,N_p}^{k^{(N_m)}_l}
    \end{pmatrix}
    \begin{pmatrix}
    f_1 \\
    \vdots\\
    f_{N_p}
    \end{pmatrix}
    = 
    \begin{pmatrix}
    M_{k^{(1)}_1,\ldots,k^{(1)}_d} \\
    \vdots\\
    M_{k^{(N_m)}_1,\ldots,k^{(N_m)}_d}
    \end{pmatrix}.\label{eq:f-linear}
\end{equation}
Here $k^{(j)}_1,\ldots,k^{(j)}_d$, $j=1,\ldots,N_m$ are the multi-indices corresponding to the orders of the given moments of the distribution function, that is, the moment constraints.
We can re-write this system succinctly as
\begin{equation}
    \mathbf{V} \mathbf{f} = \mathbf{M}.
\end{equation}
A solution $f$ of the system of constraints~(\ref{eq:f-linear}) that is sparse can be found by solving the regularized system of equations using the $L_0$ semi-norm of $f$:
\begin{equation}
    \min_{\substack{f_{i},\:1\leq i \leq N_p}} ||\mathbf{M} - \mathbf{V} \mathbf{f}||_2^2 + \lambda_0 ||\mathbf{f}||_0.\label{eq:optimization-linear}
\end{equation}
However, the problem~(\ref{eq:optimization-linear}) is computationally expensive due to use of the semi-norm $||\mathbf{f}||_0$. The more frequent approach of enforcing sparsity by regularization via the $L_1$ norm of the solution vector is not applicable due to the fact that $||\mathbf{f}||_1$ is simply the number density, a conserved quantity. In addition, the use of a fixed set of velocities in ansatz~(\ref{eq:f-expansion}) invariably leads to bias in the moments of the distribution function~\cite{dimarco2014numerical}. Finally, a non-negative solution is sought, as negative values of $f_i$ are unphysical, a constraint which needs to be incorporated into the solution of~(\ref{eq:f-linear}). Non-negative least squares approaches~\cite{slawski2013non,foucart2014sparse,oblapenko2024non} can be used to find a sparse non-negative solution of~(\ref{eq:f-linear}), but also require some choice of the positions of the Diracs, and may not necessarily converge. They also do not allow for a constrained optimization of some additional functional, like entropy.
Therefore, in the present work, we focus on optimizing not the coefficients $f_i$, but the locations of the Dirac deltas $v_{l,i}$.
Denoting the zero-order moment $M_{0,\ldots,0}$ (which corresponds to the number density) as $\rho$, we simply take the coefficients $f_i$ to be equal to $\rho/N_p$. This ensures the non-negativity of $f$ and conservation of the zero-order moment. We can assume without loss of generality that $\rho=1$, as otherwise we can always re-scale $f$ and $M_{k_1,\ldots,k_d}$ accordingly.

The representation~(\ref{eq:f-expansion}) is not sparse, as all the expansion functions have a constant non-zero coefficient $f_i=1/N_p$ associated with them. However, in case we have that $(v_{1,i_1},\ldots,v_{d,i_1})=(v_{1,i_2},\ldots,v_{d,i_2})$ for some $i_1 \neq i_2$, we can re-write expansion~(\ref{eq:f-expansion}) as
\begin{equation}
    f(\mathbf{v}) = \sum_{\substack{i=1\\ i \notin \{i_1,i_2\} }  }^{N_p} \frac{\rho}{N_p} \prod_{l=1}^{d} \delta(v_l - v_{l,i}) + 2 \frac{\rho}{N_p}\prod_{l=1}^{d} \delta(v_l - v_{l,i_1}).\label{eq:f-expansion2}
\end{equation}
This can be easily extended to the case of more coincident Dirac deltas:
\begin{equation}
    f(\mathbf{v}) = \sum_{i=1}^{N_s} \mathrm{card}(D_i)\frac{\rho}{N_p} \prod_{l=1}^{d} \delta(v_l - v_{l,D_i}).\label{eq:f-expansion3}
\end{equation}
Here $N_s$ is the number of distinct Dirac deltas appearing in~(\ref{eq:f-expansion}) and $\mathrm{card}(D_i)$ is the cardinality of the subset of identical Dirac deltas $D_i$ with position $(v_{1,D_i},\ldots,v_{y,D_i})$ (i.e. how many Dirac deltas are located at $(v_{1,D_i},\ldots,v_{l,D_i})$).

With this closure, the system of the transport equations~(\ref{eq:transport}) for the moments of order up to $N$ can be written as
\begin{align}
    & \frac{\partial M_{k_1,\ldots,k_d}}{\partial t} + \nabla \cdot \mathcal{F}_{k_1,\ldots,k_d} = 0,   & \hspace{-5em} 0 \leq \sum_{l=1}^d k_l \leq N-1,\nonumber \\
    &\frac{\partial M_{k_1,\ldots,k_d}}{\partial t} + \frac{\rho}{N_p} \nabla \cdot \mathcal{F}^{\mathrm{closure}}_{k_1,\ldots,k_d} = 0, &\sum_{l=1}^d k_l = N.
\end{align}
Here the flux $\mathcal{F}^{\mathrm{closure}}_{k_1,\ldots,k_d}$ is given through the ansatz~(\ref{eq:f-expansion3}) as
\begin{equation}
    \mathcal{F}^{\mathrm{closure}}_{k_1,\ldots,k_d} = \left(\sum_{i=1}^{N_s} \mathrm{card}(D_i)v_{1,D_i}^{k_1+1} \cdot \ldots \cdot v_{d,D_i}^{k_d},\ldots,\sum_{i=1}^{N_s} \mathrm{card}(D_i) v_{1,D_i}^{k_1} \cdot \ldots \cdot v_{d,D_i}^{k_d+1}\right).
\end{equation}

\subsection{Sparsity-promoting optimization problem}
We now pose the optimization problem we will use to determine the values $v_{l,i}$, $i=1,\ldots,N_p$, $l=1,\ldots,d$. We minimize combination of two functions: an objective function $S: \mathbb{R}^{dN_p+1} \to \mathbb{R}$, the minimization of which is expected to improve the ``quality'' of the reconstruction of the underlying unknown $f$, and a regularization term $R: \mathbb{R}^{dN_p} \to \mathbb{R}$, that promotes sparsity in the sense described above, by ``pushing together'' the Dirac deltas so that the solution of the problem (\ref{eq:optimization}) contains multiple Diracs located at the same point.

Thus, we write the optimization problem as:
\begin{equation}
  \begin{aligned}
      \min_{\substack{v_{l,i},\:1\leq i \leq N_p,\: 1 \leq l \leq d}} \quad &  S(\rho,v_{1,1},\ldots,v_{1,{N_p}},\ldots,v_{d,1},\ldots,v_{d,{N_p}})  + \lambda R(v_{1,1},\ldots,v_{1,{N_p}},\ldots,v_{d,1},\ldots,v_{d,{N_p}})\\
  \textrm{s.t.}  \quad & \sum_i \frac{\rho}{N_p}  \prod_{l=1}^{d} v_{l,i}^{k_l} = M_{k_1,\ldots,k_d}, \: 1 \leq \sum_{l=1}^d k_l \leq N,\\
  \textrm{s.t.} \quad & v_{l,min} \le v_{l,i} \le v_{l,max}, \: 1\leq i \leq N_p, \:  1 \leq l \leq d.
  \end{aligned}\label{eq:optimization}
\end{equation}
This is a non-linearly constrained optimization problem with the objective function being the sum of the two functions $S$ and $R$. The parameter $\lambda$ defines the strength of the regularization --- the larger its value, the more weight is given to finding a sparse solution. The values $v_{l,min}$, $v_{l,max}$ are the prescribed lower and upper bounds of the values of the solution values for the dimension $l$. It should again be stressed that the complexity of the problem (\ref{eq:optimization}) is linear with respect to both the number $N_p$ of Diracs used, as well as the dimension $d$ of the domain of $f$.

Once a solution of (\ref{eq:optimization}) has been found, we need to reduce it to the form~(\ref{eq:f-expansion3}) by finding which Dirac deltas coincide. As we are dealing with numerical optimization, we consider ``identical'' in an approximate sense: given a distance metric $\mathcal{D}: \mathbb{R}^d \times \mathbb{R}^d \to \mathbb{R}$ and a tolerance $r$, we assume that if 
\begin{equation}
    \mathcal{D}((v_{1,i_1},\ldots,v_{d,i_1}), (v_{1,i_2},\ldots,v_{d,i_2})) \leq r, \: i_1\neq i_2,
\end{equation}
we can replace
\begin{equation}
    \frac{\rho}{N_p} \prod_{l=1}^{d} \delta(v_l - v_{l,i_1}) + \frac{\rho}{N_p} \prod_{l=1}^{d} \delta(v_l - v_{l,i_2}) 
\end{equation}
by
\begin{equation}
    \frac{2\rho}{N_p} \prod_{l=1}^{d} \delta(v_l - v_{l,n}),
\end{equation}
where $(v_{1,n},\ldots,v_{d,n}) \in \mathbb{R}^d$ is a representative point satisfying
\begin{equation}
    \mathcal{D}((v_{1,i_1},\ldots,v_{d,i_1}), (v_{1,i_n},\ldots,v_{d,i_n})) \leq r, \: \mathcal{D}((v_{1,i_2},\ldots,v_{d,i_2}), (v_{1,i_n},\ldots,v_{d,i_n})) \leq r.
\end{equation}
To find expansion terms that are identical up to a tolerance $r$, we construct a connectivity graph of points $\{(v_{1,i},\ldots,v_{d,i})\}_{i=1}^{N_p}$, assuming that points are connected if the distance (as defined by the metric $\mathcal{D}$) between them is smaller than $r$. Once constructed, one can use a depth or breadth-first search to find the connected components in the graph, which correspond to the subset of identical Dirac deltas $I_i$ appearing in~(\ref{eq:f-expansion3}). As we need to compute the distances between all the Dirac delta functions in order to construct the connectivity graph, and each distance computation involves $\mathcal{O}\left(d\right)$ operations, this step carries an overall computational cost of $\mathcal{O}\left((d N_p)^2\right)$.

\subsection{Regularization}
In the current work, we use the Euclidean distance metric to construct the connectivity graph and find subsets of close points:
\begin{equation}
    \mathcal{D}((v_{1,i_1},\ldots,v_{d,i_1}), (v_{1,i_2},\ldots,v_{d,i_2})) = \sqrt{\sum_{l=1}^{d}(v_{l,i_1} - v_{l,i_2})^2}.
\end{equation}
Since we are optimizing for this metric, i.e. we are interested in solutions that have as little as possible distinct expansion functions w.r.t. the metric $\mathcal{D}$, it makes sense to define the regularization term $R$ as
\begin{equation}
    R=R_2(v_{1,1},\ldots,v_{1,{N_p}},\ldots,v_{d,1},\ldots,v_{d,{N_p}}) = \frac{1}{N_p^2}\sum_{i \neq j}  \sqrt{\sum_{l=1}^{d}(v_{l,i} - v_{l,j})^2}.\label{eq:sparsity:euclid}
\end{equation}
Minimization of $R$ clusters the Dirac deltas close together, thus promoting a sparse solution, in the sense described above.

An alternative definition of $R$, also considered in the present work, is based on the Manhattan distance:
\begin{equation}
    R=R_1(v_{1,1},\ldots,v_{1,{N_p}},\ldots,v_{d,1},\ldots,v_{d,{N_p}}) = \frac{1}{N_p^2}\sum_{i \neq j} \sum_{l=1}^{d} |v_{l,i} - v_{l,j}|.\label{eq:sparsity:manhattan}
\end{equation}
This is somewhat less expensive to evaluate than the regularization term given by~(\ref{eq:sparsity:euclid}), however, this does not directly optimize the sparsity induced by the Euclidean distance metric. Nevertheless, we consider both possibilities for the regularization term and assess their impact on the sparsity and accuracy of the solution in the section on numerical results. Similar to the construction of the connectivity graph, computing the regularization term has a computational cost of $\mathcal{O}\left((d N_p)^2\right)$.

\subsection{Entropy as objective function}
The last remaining thing to specify is the function $S$. Motivated by the maximum entropy approach of classical moment methods, where entropy minimization is used to obtain a unique non-negative solution to the moment problem that ensures strict hyperbolicity~\cite{levermore1997entropy} of the associated transport equations, we are interested in minimizing the kinetic theory entropy, defined as
\begin{equation}
    \mathcal{S}(f) = \int f \log f  \mathrm{d}\mathbf{v}.
\end{equation}
However, we cannot compute $\mathcal{S}(f)$ for our ansatz~(\ref{eq:f-expansion}) due to the expansion in Dirac delta functions~\cite{martin2016octree}.

Therefore, we use an approximation of $\mathcal{S}(f)$ via histograms. We define a histogram bin $I_{b_1,\ldots,b_d}$, $b_l \in \mathbb{N},l=1,\ldots,d$ as
\begin{multline}
    I_{b_1,\ldots,b_d} = \left\{(v_1,\ldots,v_d) | v_{l,min} + (b_l-1)\Delta v_l \leq v_l \leq v_{l,min} + b_l\Delta v_l,\: l=1,\ldots,d \right\},\\
    1\leq b_l \leq N_{h,l},\: l=1,\ldots,d.
\end{multline}
where $\Delta v_l = (v_{l,max} - v_{l,min})/N_{h,l}$, $N_{h,l}$ is the number of histograms in the $l$-th dimension.
We then define the counting function $\mathcal{C}_{b_1,\ldots,b_d}$ that computes the number of particles in each bin as
\begin{equation}
    \mathcal{C}_{b_1,\ldots,b_d} = \sum_{k=1}^{N_p}\chi_{b_1,\ldots,b_d}(v_{1,k},\ldots,v_{d,k}).
\end{equation}
Here $\chi_{b_1,\ldots,b_d}: \mathbb{R}^d \to \{0,1\}$ is the indicator function for bin $I_{b_1,\ldots,b_d}$.

\begin{figure}[h]
  \centering
    \begin{tikzpicture}
    \def\cellsize{1.5}

    \foreach \row in {1,2,3} {
      \foreach \col in {1,2,3} {
        \pgfmathtruncatemacro{\x}{\col - 1}
        \pgfmathtruncatemacro{\y}{3 - \row}
        \pgfmathsetmacro{\Xpos}{\x * \cellsize}
        \pgfmathsetmacro{\Ypos}{\y * \cellsize}
        \draw (\Xpos,\Ypos) rectangle ++(\cellsize,\cellsize);
        \pgfmathtruncatemacro{\rowm}{4 - \row}
        \pgfmathtruncatemacro{\colm}{\col}
        \ifnum\row=1
          \ifnum\col=2
           \fill (\Xpos + 0.25*\cellsize, \Ypos + 0.1*\cellsize) circle (2.5pt);
           
           \fill (\Xpos + 0.45*\cellsize, \Ypos + 0.4*\cellsize) circle (2.5pt);
           
           \fill (\Xpos + 0.75*\cellsize, \Ypos + 0.2*\cellsize) circle (2.5pt);
             \node at (\Xpos + 0.45*\cellsize, \Ypos + 0.7*\cellsize) [align=left] {$I_{\row, \col}$\\$\mathcal{C}_{\rowm, \colm}=3$};
          \else
            \node at (\Xpos + 0.45*\cellsize, \Ypos + 0.7*\cellsize) [align=left]{$I_{\row, \col}$\\$\mathcal{C}_{\rowm, \colm}=0$};
          \fi
        \else\ifnum\row=3
          \ifnum\col=1
            
              \fill (\Xpos + 0.65*\cellsize, \Ypos + 0.73*\cellsize) circle (2.5pt);
              \fill (\Xpos + 0.15*\cellsize, \Ypos + 0.23*\cellsize) circle (2.5pt);
        
             \node at (\Xpos + 0.45*\cellsize, \Ypos + 0.7*\cellsize) [align=left] {$I_{\row, \col}$\\$\mathcal{C}_{\rowm, \colm}=2$};
          \else
            \node at (\Xpos + 0.45*\cellsize, \Ypos + 0.7*\cellsize) [align=left]{$I_{\row, \col}$\\$\mathcal{C}_{\rowm, \colm}=0$};
          \fi
        \else
        
          \node at (\Xpos + 0.45*\cellsize, \Ypos + 0.7*\cellsize) [align=left] {$I_{\row, \col}$\\$\mathcal{C}_{\row, \col}=0$};
        \fi
        \fi
      }
    }

    \node[anchor=south east] at (0, 3*\cellsize) {$(v_{1,min},v_{2,max})$};
    \node[anchor=south west] at (3*\cellsize, 3*\cellsize) {$(v_{1,max},v_{2,max})$};
    \node[anchor=north east] at (0, 0) {$(v_{1,min},v_{2,min})$};
    \node[anchor=north west] at (3*\cellsize, 0) {$(v_{1,max},v_{2,min})$};
  \end{tikzpicture}
  


  \caption{Example of the histogram binning of the representation of the distribution function via Diracs (\ref{eq:f-expansion}) (denoted by the black dots) in a 2-dimensional setting with 3 bins in each velocity direction. }
  \label{fig:grid-diagram}
\end{figure}
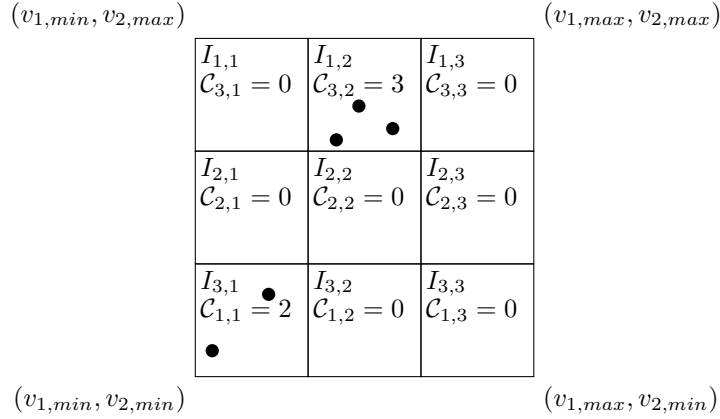
For illustrative purposes, a schematic of the binning and computed values of the counting function is shown on Figure~\ref{fig:grid-diagram} in a 2-dimensional setting for a 3$\times$3 grid with $N_p=5$ Diracs.

We can then define the histogram-approximated entropy as
\begin{equation}
     S(\rho,v_{1,1},\ldots,v_{1,{N_p}},\ldots,v_{d,1},\ldots,v_{d,{N_p}}) =  \frac{\rho}{N_p}  \sum_{b_1=1}^{N_{h,1}} \ldots \sum_{b_d=1}^{N_{h,d}} \mathcal{C}_{b_1,\ldots,b_d} \log \left( \frac{\rho}{N_p} \mathcal{C}_{b_1,\ldots,b_d} \right) \mu(I_{b_1,\ldots,b_d}).\label{eq:s-histogram}
\end{equation}
Here $\mu(I_{b_1,\ldots,b_d}):\mathbb{R}^d \to \mathbb{R}$ is the measure of bin $I_{b_1,\ldots,b_d}$. With this definition of the approximated entropy and the sparsity promoting terms (either~(\ref{eq:sparsity:euclid}) or~(\ref{eq:sparsity:manhattan})), the optimization problem~(\ref{eq:optimization}) is fully defined. Next, we discuss some aspects of solving~(\ref{eq:optimization}) in practice.

\subsection{Aspects of numerical implementation}

\subsubsection{Efficient computation of histogram-approximated entropy}
In order to avoid the multiple summations appearing in~(\ref{eq:s-histogram}) and having to check for empty bins which need to be skipped to avoid taking the logarithm of 0, for practical purposes the computation of $S(f)$ is performed by using a dictionary structure consisting of key-value pairs 
\begin{equation}
(b_1,\ldots,b_d) \to \sum_{k=1}^{N_p} \chi_{b_1,\ldots,b_d}(v_{1,k},\ldots,v_{d,k}).
\end{equation}
At the start, an empty dictionary is initialized, and then a single pass over the $N_p$ Dirac delta functions is performed. For each Dirac delta its corresponding bin indices $(b_1,\ldots,b_d)$ are computed (an operation with complexity of $\mathcal{O}(d)$) and used as the key. If no such key is present in the dictionary, a new element with value $1$ is added; if the key is already present, the associated stored value is increased by $1$. As the worst case performance of a dictionary using a hash map is $\mathcal{O}(N_{el})$ (where $N_{el}$ is the number of elements), the overall computational complexity of computing the histogram-approximated entropy is no worse than $\mathcal{O}(d N_p)$.
As it is reasonable to assume that $N_p \ll N_{h,1} \times \ldots \times N_{h,d}$, the procedure is significantly more efficient than computing and storing the values of the indicator functions of all the bins.

A second hash map can be used to store the measures of the non-empty bins $\mu(I_{b_1,\ldots,b_d})$, or, in the case of uniform bins, one can simply compute $\mu(I_{b_1,\ldots,b_d})$ when required.

\subsubsection{Entropy scaling}
In order to ensure a more similar order of magnitude of the values of $S$ and $R$ appearing in~(\ref{eq:optimization}), a scaling is used in the computation of $S$.
For the (non-approximated) entropy $\mathcal{S}(f)$ we have that
\begin{equation}
     \mathcal{S}\left(\frac{f}{C}\right) = \int \frac{f}{C} \log \frac{f}{C} dv  = \frac{1}{C} \left(\int f \log f dv -  \int f \log C dv \right) = \frac{\mathcal{S}(f)}{C} - \rho \log C,
\end{equation}
where $C > 0$ is a scaling factor.

For the histogram-approximated entropy we can equivalently write
\begin{equation}
     S(\rho/C,v_{1,1},\ldots,v_{1,{N_p}},\ldots,v_{d,1},\ldots,v_{d,{N_p}}) = \frac{S(\ldots)}{C} - \rho \log C.\label{eq:final-s-histogram}
\end{equation}

In the current work the expression~(\ref{eq:final-s-histogram}) is the one used in the optimization problem~(\ref{eq:optimization}), with $C$ computed as
\begin{equation}
     C =  \sum_{b_1=1}^{N_{h,1}} \ldots \sum_{b_d=1}^{N_{h,d}} \mathcal{C}_{b_1,\ldots,b_d}  \mu(I_{ij}).
\end{equation}
This scaling constant corresponds to the histrogram-based approximation of the total number density $\rho$ and can be easily computed based on the quantities already being computed for the approximate entropy $S$.

\subsubsection{Numerical optimization}
It should be noted that the histogram-based approximation of entropy~(\ref{eq:s-histogram}) is a piece-wise constant function. That means that any optimization approach that relies on derivative information is bound to be unsuccessful, as the partial derivatives over $v_{l,k}$ are either 0 or undefined. This fact, coupled with the non-linear moment constraints, means that relatively few optimization codes are able to solve the problem~(\ref{eq:optimization}) with the histogram-based objective function~(\ref{eq:s-histogram}). In the present work, the COBYLA (``constrained optimization by linear approximation'') algorithm~\cite{powell1994direct} is used to solve~(\ref{eq:optimization}).

It should be noted that even without the difficulties posed by the piece-wise constant function (\ref{eq:s-histogram}), the optimization is already non-trivial in case the $L_1$ norm-based regularization term $R_1$ is used. Nevertheless, the numerical results presented further show that the overall approach is valid and does indeed result in a sparse solution that satisfies the moment constraints.


\section{Numerical results}
We now investigate the performance of the proposed optimization-based approach in a two-dimensional scenario. To that extent, we consider the following standard approach: we define a hidden truth distribution $f^{\mathrm{hidden}}(v_1,v_2)$, compute its moments 
 $\{M_{k_1,k_2}^{\mathrm{hidden}}\}_{k_1+k_2\leq N}$ via analytical or numerical integration, and use them as constraints in the optimization problem~(\ref{eq:optimization}). We then use the computed $f$ to predict  $\{M_{k_1,k_2}\}_{k_1+k_2\leq N+1}$ and compute the overall error in the predicted moments:
 \begin{equation}
      E^{N+1}_{\mathrm{tot}} = \sum_{k_1+k_2= N+1} \left| M_{k_1,k_2}^{\mathrm{hidden}} - M_{k_1,k_2} \right|.
 \end{equation}
 We consider the absolute and not relative error as the metric, as we look at several distributions where some of the moments $M_{k_1,k_2}^{\mathrm{hidden}}$ have a value of 0, and a relative error would not be meaningful.
As these hidden truth distributions we consider several distributions arising from kinetic theory: a double-Dirac distribution, a Maxwell--Boltzmann distribution, a 2-dimensional Druyvesteyn distribution, and a bi-modal distribution (mixture of 2 Maxwell--Boltzmann distributions). We assume that we are working in scaled variables, so that the temperature, molecular mass, and number density do not appear in the expressions for the velocity distribution functions, except for the bimodal distribution case, where the scaled temperatures show up. We consider conservation of all moments of total order up to 3 (corresponding to a total of 10 conserved moment values) and conservation of all moments of order up to 4 (corresponding to a total of 15 conserved moment values).

We choose the lower and upper velocity bounds both for $v_1$ and $v_2$ to be -4 and 4, correspondingly; the same lower and upper bounds are used to discretize the domain into histogram bins used to approximate the entropy.

 We vary the following numerical parameters:
 \begin{enumerate}
     \item Choice of sparsity-promoting regularizer $R$: either $R_2$, as defined by (\ref{eq:sparsity:euclid}), or $R_1$, as defined by (\ref{eq:sparsity:manhattan})
     \item Value of regularization parameter $\lambda$: we consider the following values: $\lambda \in \{ 10^{-3}, 10^{-2}, 10^{-1}, 1, 10^1, 10^2, 2.5 \times 10^2, 5\times 10^2, 7.5 \times 10^2, 10^3, 5 \times 10^3, 10^4, 5 \times 10^4, 10^5, 10^6 \}$
     \item Number of histogram bins: we consider $N_{h,1} = N_{h,2} \in \{10, 20, 50 \}$
 \end{enumerate}
In all the results presented, the number of starting Diracs was taken to as $N_p=30$. For small values of $\lambda$, the regularization term has little impact on value of the objective function, and it is expected that the solution will mostly be determined by the optimization of the entropy approximation; conversely, for high values of $\lambda$ it is expected that the sparseness of the solution will be the main optimization target. Therefore, after a certain value of $\lambda$, increasing $\lambda$ further will not affect the solution.

The software was written in Julia with the use of the PRIMA library~\cite{Zhang_2023} which implements the COBYLA algorithm~\cite{powell1994direct}. The code has been made publicly available on Github~\cite{oblapenko2025reprorepo}. To improve convergence, a two-step optimization approach was used. First, a set of values $\{(v_1, v_2)\}$ is found that satisfies the given moment constraints, without minimizing any objective function. This solution is then used as the initial guess of the solution of the full optimization problem~(\ref{eq:optimization}). The tolerance $r$ used to define the distance at which the Diracs are assumed to be identical was taken to be $10^{-3}$.

\subsection{Alternative approaches to the moment closure problem}
To better assess the performance of the proposed algorithm, we consider other approaches against which we compare the quality of the solutions obtained via solving~(\ref{eq:optimization}): 
\begin{enumerate}
    \item Solving (\ref{eq:optimization}) with a constant objective function $S$ and no regularization, this corresponds to simply solving for the moment constraints without any further optimization of the positions of the Diracs.
    \item Using a Grad expansion to represent the VDF~\cite{grad2kinetic}.
    \item Using the classical maximum entropy approach~\cite{levermore1997entropy}, that is, the function $f$ is assumed to be a smooth function governed by parameters found by solving a numerical optimization problem.
\end{enumerate}

We now briefly describe the Grad and classical maximum entropy approaches used in the present work.
\subsubsection{Grad's method}

The method of Grad~\cite{grad2kinetic} involves representing the unknown VDF as an equilibrium Maxwell--Boltzmann distribution multiplied by a polynomial correction.
We can write this as
\begin{equation}
    f^{\textrm{Grad}}(v_1,v_2) = \frac{\rho}{\pi T}\exp\left(-\frac{v_1^2+v_2^2}{T} \right) \sum_{k+l=0}^{k+l=N} c^G_{k,l} v_1^k v_2^l,
\end{equation}
where the number density $\rho$ of the equilibrium Maxwellian is taken to be $\rho=M_{0,0}$, and the temperature $T$ is computed as $T=(M_{2,0} + M_{0,2})/\rho$. The number of expansion terms is chosen to be equal to the number of the moment constraints, with the monomial powers $k$, $l$ chosen accordingly.
The unknown expansion coefficients $c^G_{k,l}$ are found by solving a linear system of moment constraints:
\begin{equation}
    \begin{bmatrix} 
    a_{0,0} & a_{0,1} & a_{0,2} & \ldots & a_{0,N} & a_{1,0} & \ldots \\
    a_{1,0} & a_{1,1} & a_{1,2} &  \ldots & a_{1,N} & a_{2,0} & \ldots \\
    \vdots & &       & \ddots \\
    a_{N,0} & a_{N,1} & a_{N,2} &  \ldots & a_{N,N} & a_{N+1,0} & \ldots \\
    a_{0,1} & a_{0,2} & a_{0,3} &  \ldots & a_{0,N+1} & a_{1,1} & \ldots \\
    & &  & \ddots
    \end{bmatrix}
    \cdot 
    \begin{bmatrix} 
    c_{00}  \\
    c_{01}  \\
    \vdots \\ 
    c_{0N} \\
    c_{10} \\
    c_{11} \\ 
    \vdots \\ 
    c_{N0}
    \end{bmatrix}
    = 
    \begin{bmatrix} 
    M_{00}  \\
    M_{01}  \\
    \vdots \\ 
    M_{0N} \\
    M_{10} \\
    M_{11} \\ 
    \vdots \\ 
    M_{N0}
    \end{bmatrix}.
\end{equation}
Here the coefficients $a_{i,j}$ are computed as
\begin{equation}
    a_{i,j} = \frac{\rho}{\pi T} \int v_1^i v_2^j \exp\left(-\frac{v_1^2+v_2^2}{T} \right) \mathrm{d}v_1 \mathrm{d}v_2.
\end{equation}
Whilst the Grad closure can be easily computed for any number of moments, it leads to unphysical (negative) values of the VDF and loss of hyperbolicity in regimes with strong non-equilibrium~\cite{cai2019holway}, and as such poses challenges for general kinetic transport problems.

\subsubsection{Classical maximum entropy}
The classical maximum entropy approach~\cite{levermore1997entropy} in the 2-dimensional case searches for the VDF in the form
\begin{equation}
    f(v_1,v_2) = \exp\left(-\sum_{i_1,i_2} \lambda_{i_1,i_2} v_1^{k_{i_1}} v_2^{k_{i_2}}\right),\label{eq:f-max-entropy}
\end{equation}
where the powers $k_{i_1}$, $k_{i_2}$ are taken to be the ones used to define the moment constraints, and $\lambda_{i_1,i_2}$ are the unknown coefficients.
In this case one can solve the following optimization problem to find their values:
\begin{equation}
  \begin{aligned}
      \min_{\substack{\lambda_{i_1,i_2}}} \quad &  -\int \exp\left(-\sum_{i_1,i_2}  \lambda_{i_1,i_2} v_1^{k_{i_1}} v_2^{k_{i_2}}\right) \sum_{i_1,i_2}  \lambda_{i_1,i_2} v_1^{k_{i_1}} v_2^{k_{i_2}} \mathrm{d}v_1 \mathrm{d}v_2\\
  \textrm{s.t.}  \quad &  \int \exp\left(-\sum_{i_1,i_2}  \lambda_{i_1,i_2} v_1^{k_{i_1}} v_2^{k_{i_2}}\right) v_{1}^{k_1} v_{2}^{k_2} \mathrm{d}v_1 \mathrm{d}v_2 = M_{k_1,k_2}, \: 0 \leq k_1 + k_2 \leq N.
  \end{aligned}\label{eq:optimization-lambda}
\end{equation}

In this case the entropy is a smooth function, so one can use an optimizer that can take advantage of the gradient and Hessian information. However, in the vicinity of equilibrium and the realizability boundary the problem becomes ill-conditioned and requires additional regularization to be solved~\cite{alldredge2019regularized,sadr2024wasserstein}. Such regularizations are not considered in the present work.
The numerical solution found using IPOPT via the JuMP library interface~\cite{Lubin2023}. A 2-dimensional $60^2$ quadrature was used to evaluate the integrals required for the computation of the entropy and moments after a change of variables so as to transform the integration over an infinite domain into an integration over a hypercube. The classical maximum entropy approach can only be used to predict odd-order moments using moment constraints on lower-order moments, otherwise the integrals appearing in~(\ref{eq:optimization-lambda}) diverge, although discrete-valued approaches~\cite{bohmer2020entropic,yilmaz2024nonlinear} that operate on a finite support alleviate this restriction. As we only consider the classical maximum entropy method in the present work, the results obtained with this approach are presented only for the case of $N=4$.

\subsubsection{Qualitative comparison of different closure approaches}
\begin{figure}[h]
  \centering
  \includegraphics[width=.725\textwidth]{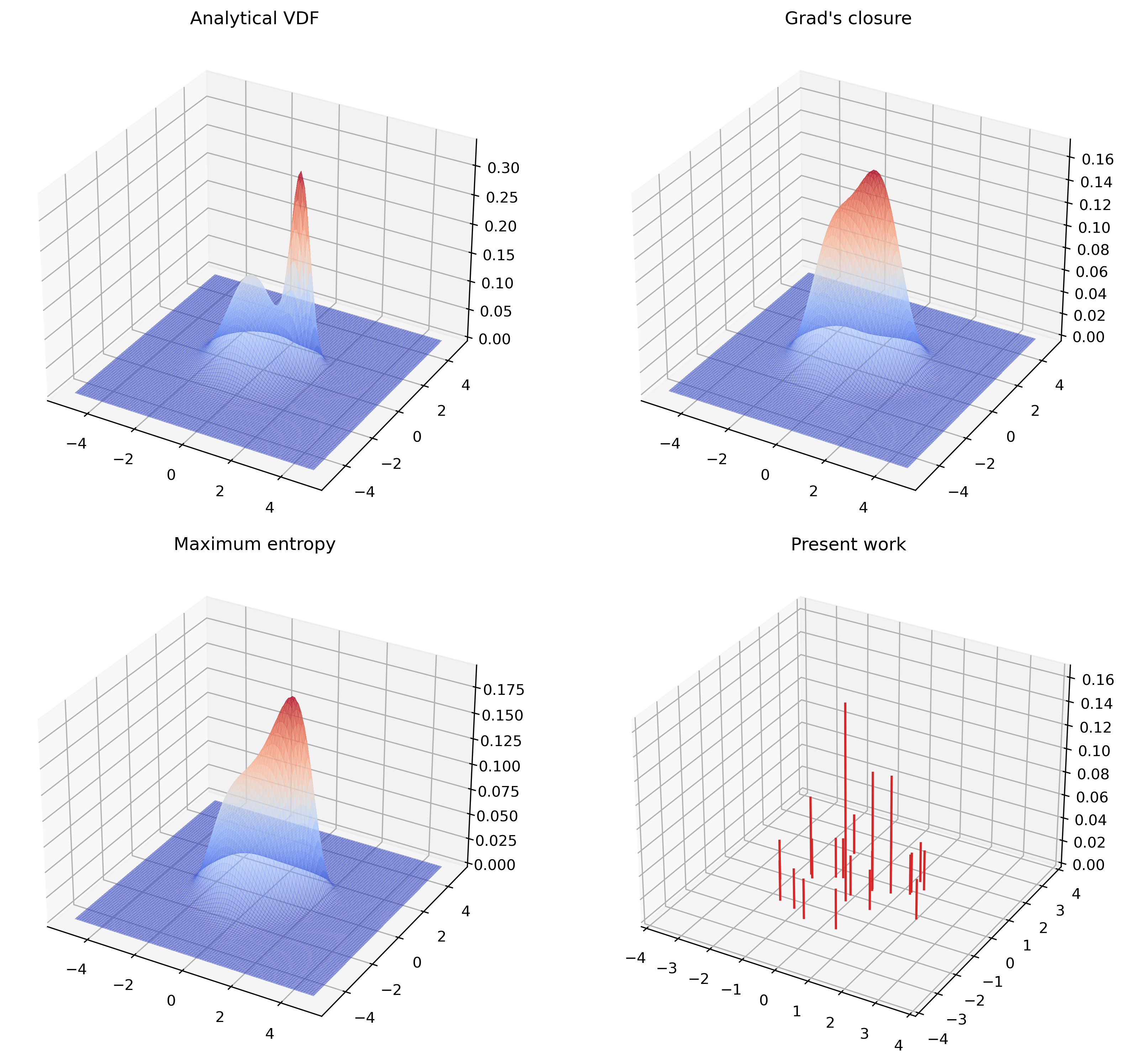}
  \caption{The analytical VDF (top left) and distribution functions obtained using different closures.}
  \label{fig:bim4-vdf}
\end{figure}
Figure~\ref{fig:bim4-vdf} shows plots of the velocity distribution function obtained using the different methods, as well as the plot of the analytical values of the hidden-truth VDF (upper left subplot) for the case of a bimodal mixture of two Maxwell--Boltzmann distributions (considered in detail in Sec.~\ref{sec:bimodal}).
The representation~(\ref{eq:f-expansion}) developed in the present work is shown in the lower right subplot, with the height of the spikes corresponding to their weights $\mathrm{card}(D_i)\frac{\rho}{N_p}$, as given by~(\ref{eq:f-expansion3}). 
The non-smooth nature of this representation a direct comparison of the VDF; however, the approximate features of the VDF are retained --- higher values of the VDF are observed around the mean velocity (taken to be $(0,0)$ in the numerical example), and the VDF rapidly decays away from the mean velocity.

\subsection{Two-Dirac distribution}
We first consider a distribution at the realizability boundary: a mixture of two Dirac delta functions:
\begin{equation}
    f(v_1,v_2) = \rho_1 \delta(v_1 - v_{1,1})\delta(v_1 - v_{2,1}) + \rho_2 \delta(v_1 - v_{1,2})\delta(v_2 - v_{2,2}).
\end{equation}
We choose $\rho_1 = 1/3$, $\rho_2 = 2/3$, $v_{1,1}=-2$, $v_{2,1}=-1$, $v_{1,2}=1$, $v_{2,2}=1/2$. Since we do not optimize for the weights associated with the Dirac deltas, the proposed method should be able to recover such a distribution as long as $N_p$ (the number of Diracs used) is divisible by the lowest common denominator of the density ratios. In the present case, $N_p=30$ and the lowest common denominator is 3; therefore, the distribution is expected to be successfully recovered by solving~(\ref{eq:optimization}).

\begin{figure}[h]
  \centering
  \includegraphics[width=.47\textwidth]{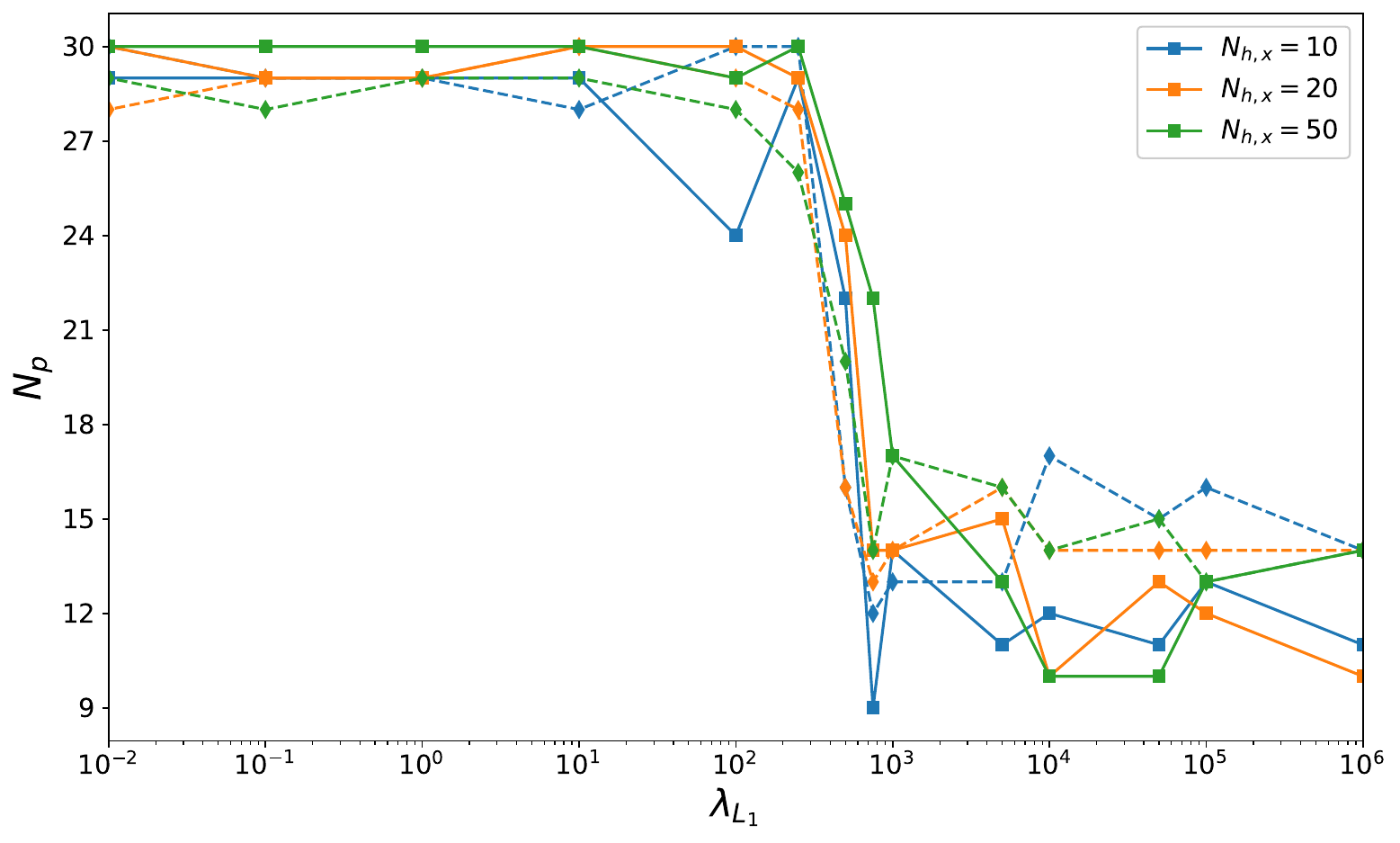}
  \includegraphics[width=.47\textwidth]{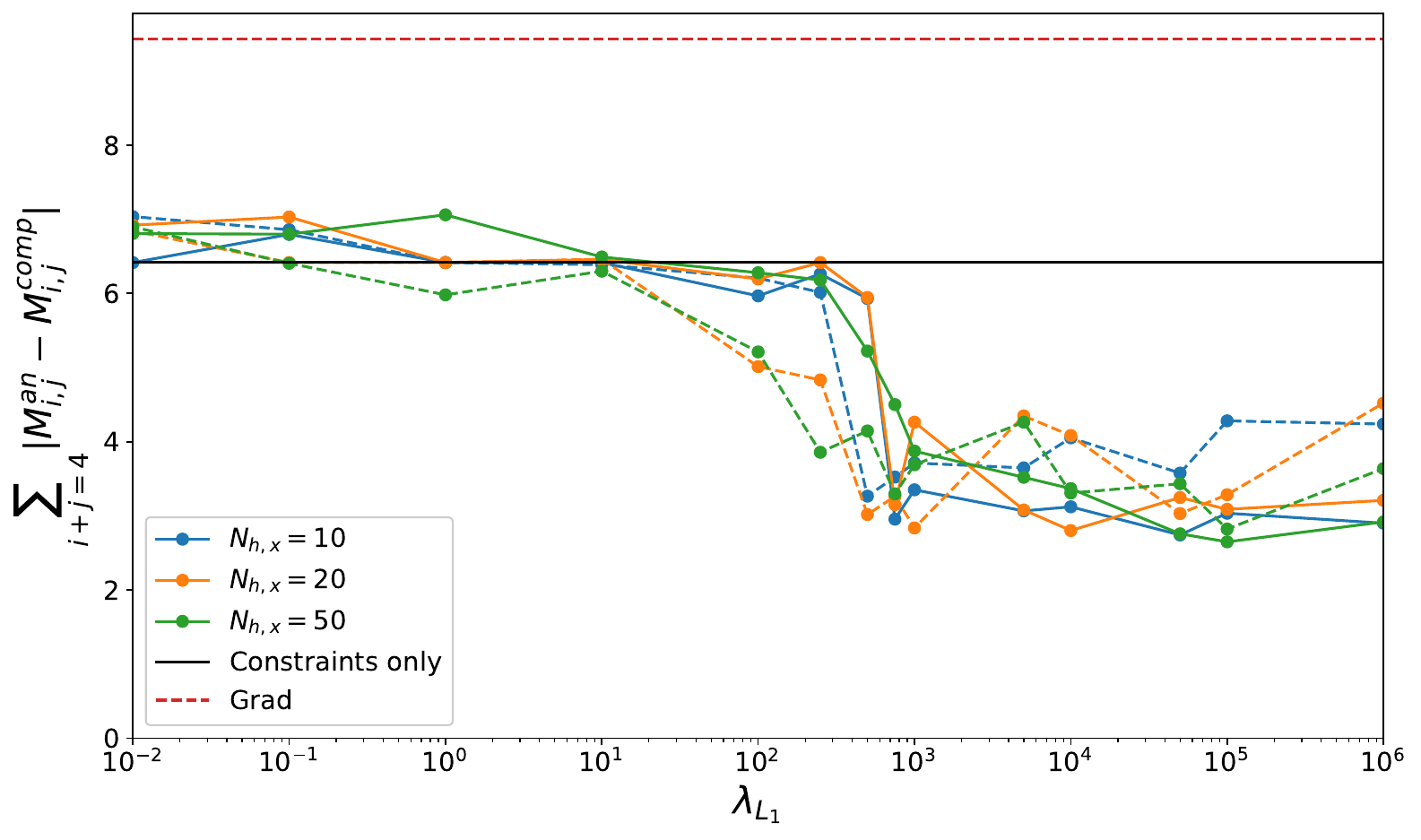}
  \caption{Number of distinct Diracs (left) and error in predicted moments of total order 4 (right) as a function of the sparsity-promoting regularization strength $\lambda_{L_1}$ for the case of $N=3$; double-Dirac distribution. Solid lines: $R_2$ regularization term, dashed lines: $R_1$ regularization term. Analytical value of $\sum_{i+j=N+1}|M_{i,j}^{an}|$ is 11.625.}
  \label{fig:dd3}
\end{figure}

\begin{figure}[h]
  \centering
  \includegraphics[width=.47\textwidth]{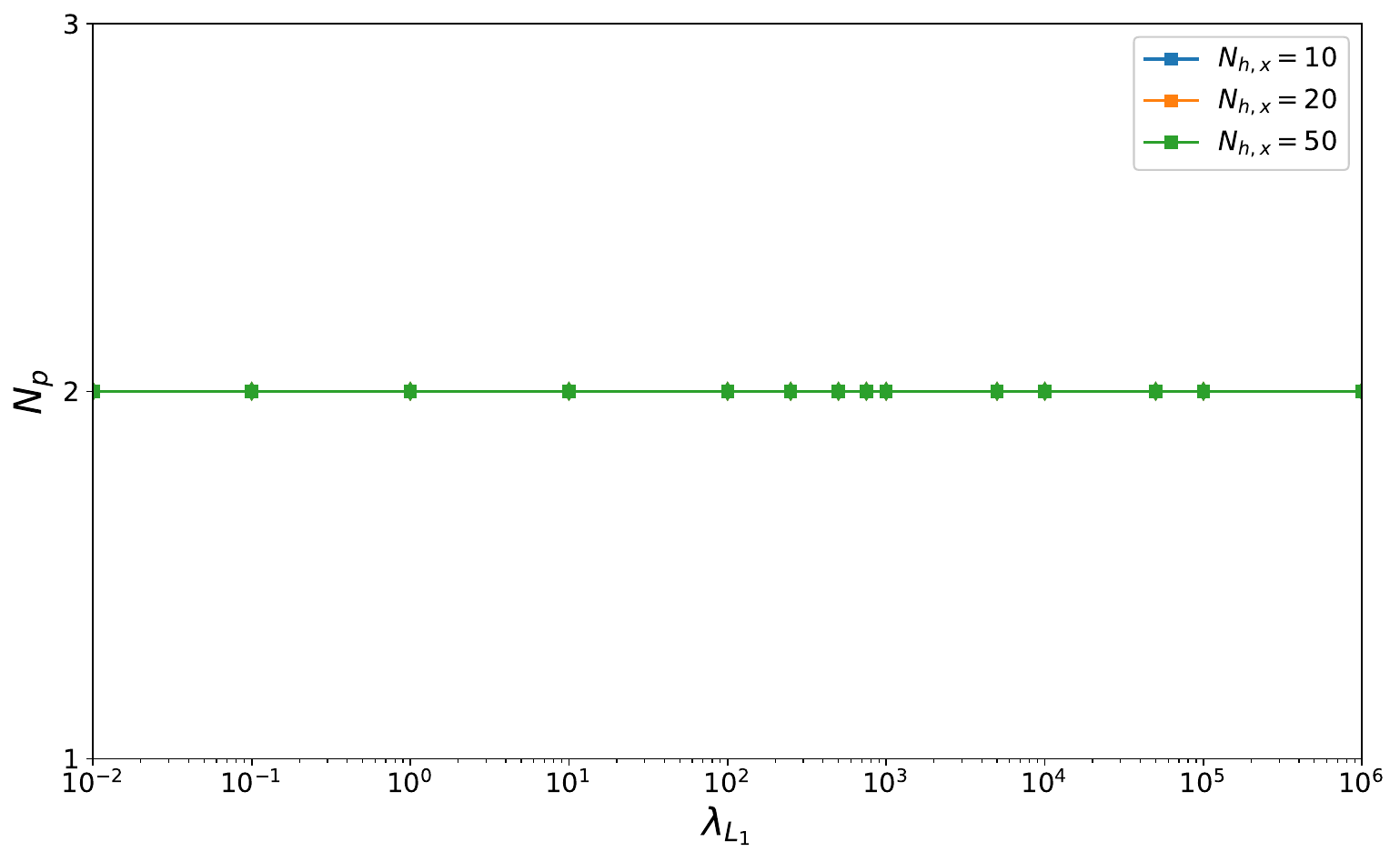}
  \includegraphics[width=.47\textwidth]{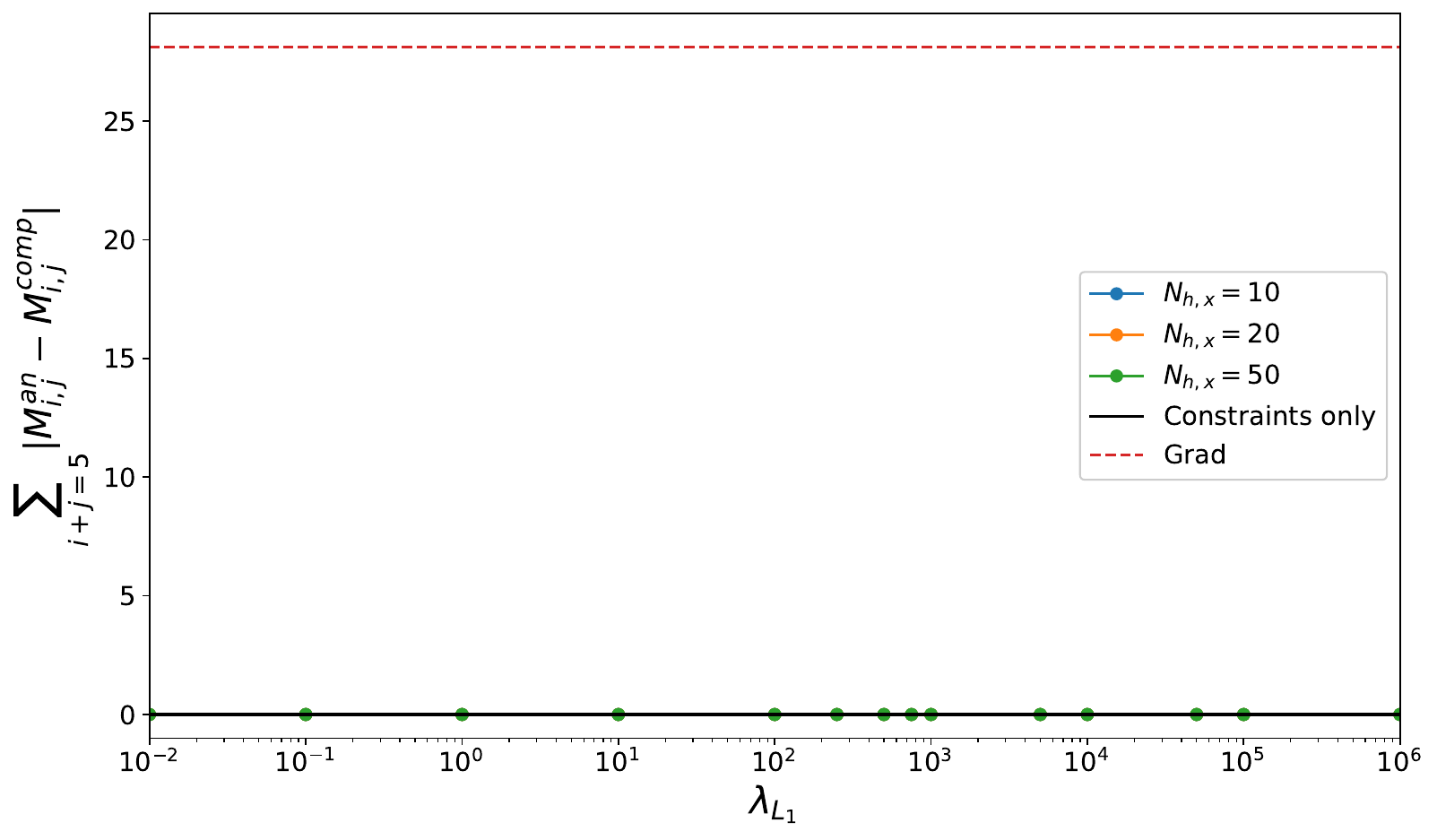}
  \caption{Number of distinct Diracs (left) and error in predicted moments of total order 5 (right) as a function of the sparsity-promoting regularization strength $\lambda_{L_1}$ for the case of $N=4$; double-Dirac distribution. Solid lines: $R_2$ regularization term, dashed lines: $R_1$ regularization term. Analytical value of $\sum_{i+j=N+1}|M_{i,j}^{an}|$ is 19.6875.}
  \label{fig:dd4}
\end{figure}

Figures~\ref{fig:dd3}--\ref{fig:dd4} show the number of distinct Diracs (left) and the error in the next predicted moments (right) for the case of constraints on all moments of total order up to 3 (Figure~\ref{fig:dd3}) and constraints on all moments of total order up to 4 (Figure~\ref{fig:dd4}). When all moments of total order up to 3 are considered, the method does not yet converge to just 2 Dirac delta functions, however, enforcing more sparsity by increasing $\lambda_1$ does lead to a lower error, as seen on~Figure~\ref{fig:dd3} (right).  The Grad closure gives significantly higher error than the proposed closure. In this case, the choice of the number of bins for the histogram approximation of entropy and the choice of the distance metric
for the sparsity-promoting regularization term has very little impact on the results.

When additional higher-order moment constraints are added, the two-Dirac underlying VDF is recovered, as evidenced by the 2 distinct Diracs obtained as a solution (Figure~\ref{fig:dd4}, left) and the zero error (Figure~\ref{fig:dd4}, right). The classical maximum entropy closure problem~(\ref{eq:optimization-lambda}) does not converge in this case due to lack of regularization, whereas Grad's method produces a very high error.

\subsection{Maxwell--Boltzmann distribution}
Next, as the hidden truth distribution, we consider the 2-dimensional Maxwell--Boltzmann distribution with a scaled temperature of 1:
\begin{equation}
    f^{\mathrm{hidden}}(v_1,v_2) =  \frac{1}{\pi}\exp\left(-v_1^2-v_2^2 \right).
\end{equation}

\begin{figure}[h]
  \centering
  \includegraphics[width=.47\textwidth]{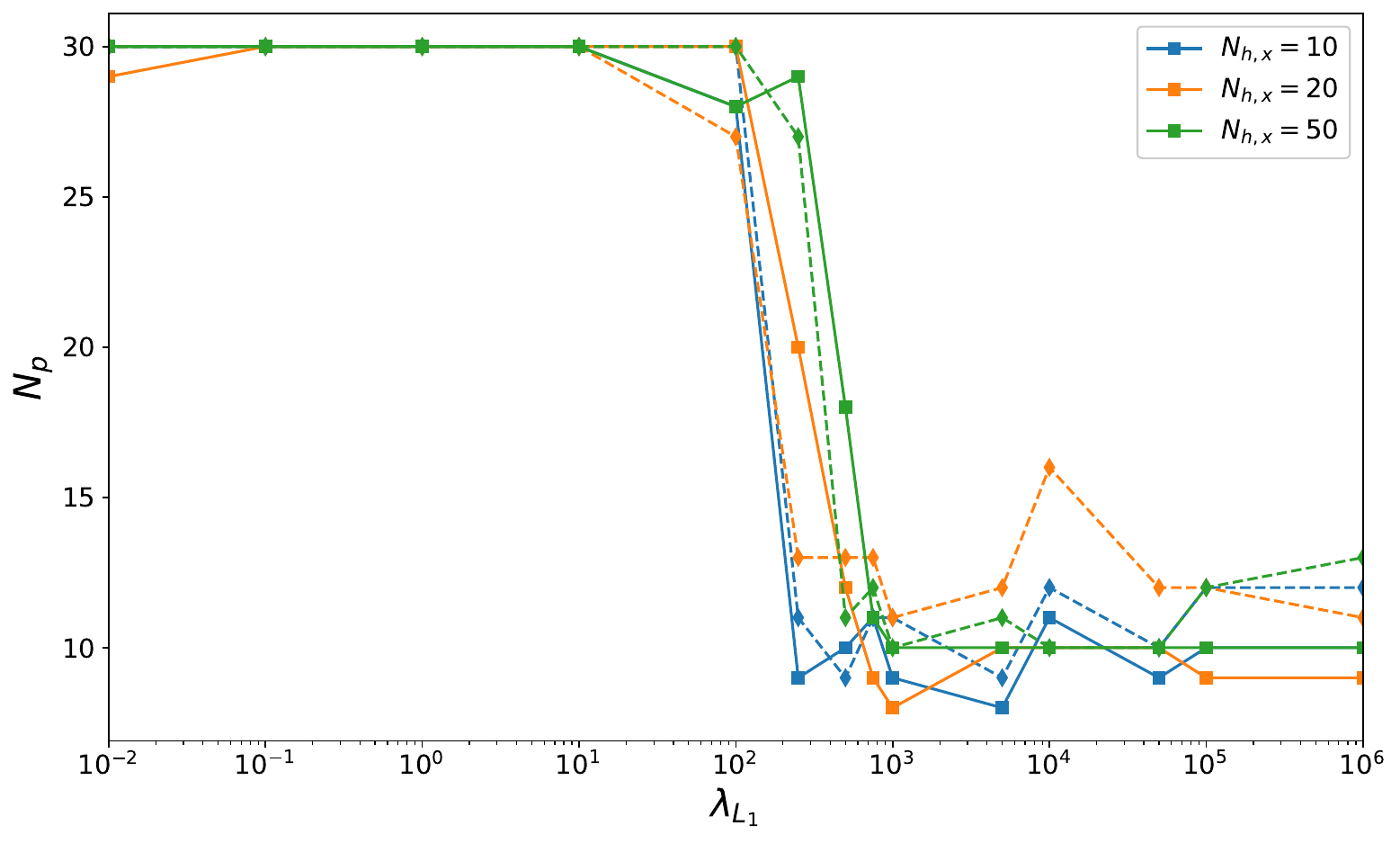}
  \includegraphics[width=.47\textwidth]{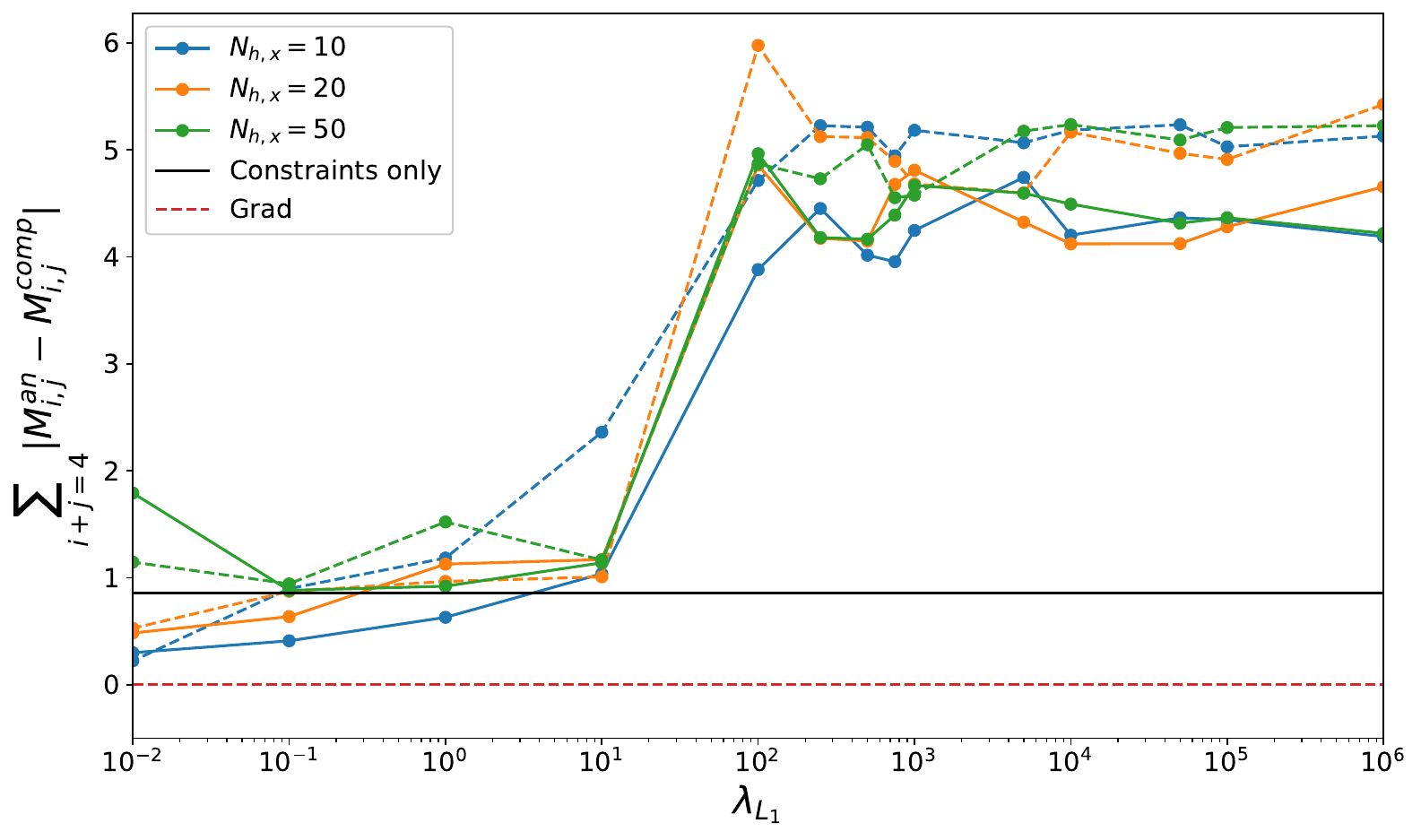}
  \caption{Number of distinct Diracs (left) and error in predicted moments of total order 4 (right) as a function of the sparsity-promoting regularization strength $\lambda_{L_1}$ for the case of $N=3$; Maxwell--Boltzmann distribution. Solid lines: $R_2$ regularization term, dashed lines: $R_1$ regularization term. Analytical value of $\sum_{i+j=N+1}|M_{i,j}^{an}|$ is 1.75.}
  \label{fig:mb3}
\end{figure}

\begin{figure}[h]
  \centering
  \includegraphics[width=.47\textwidth]{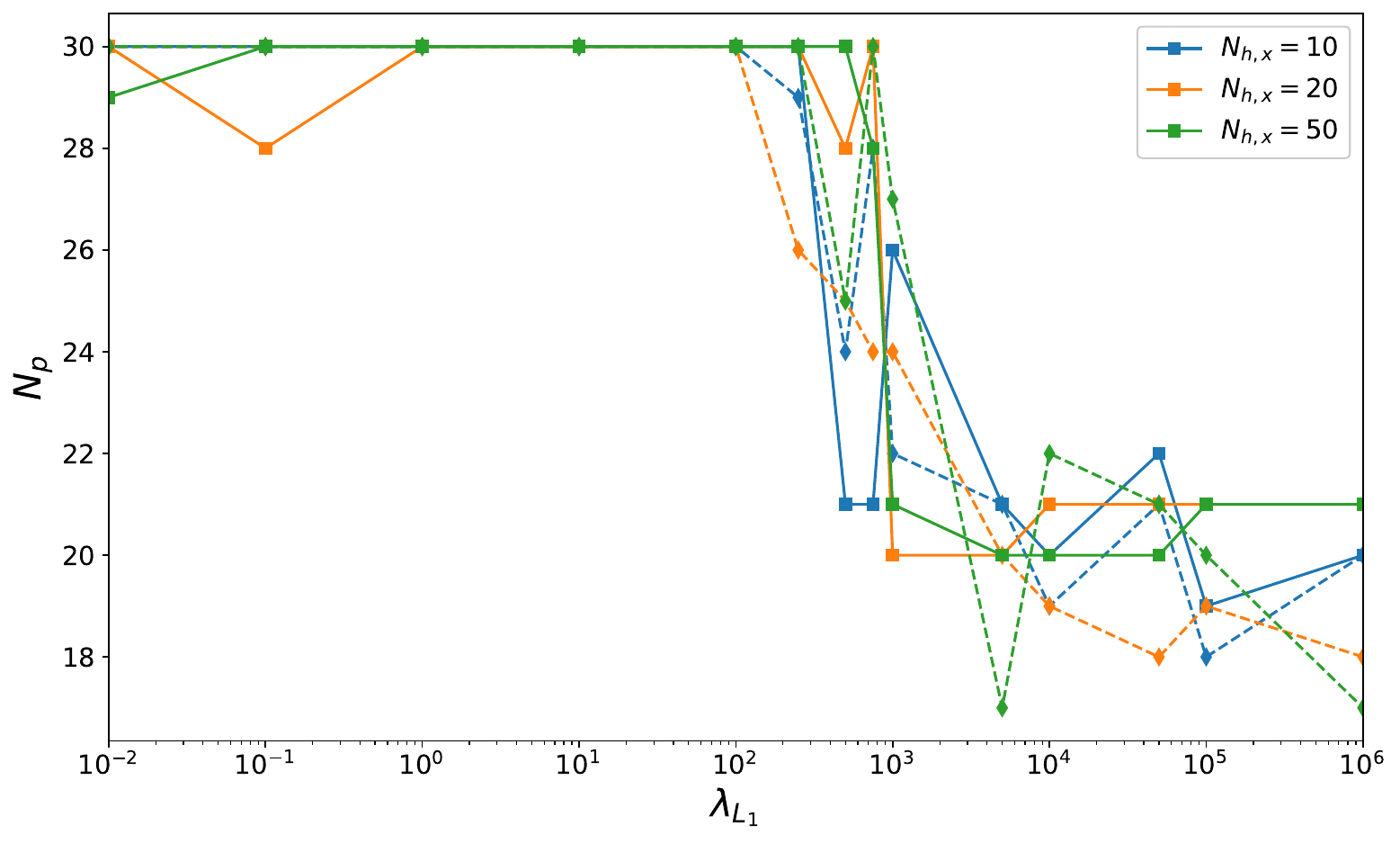}
  \includegraphics[width=.47\textwidth]{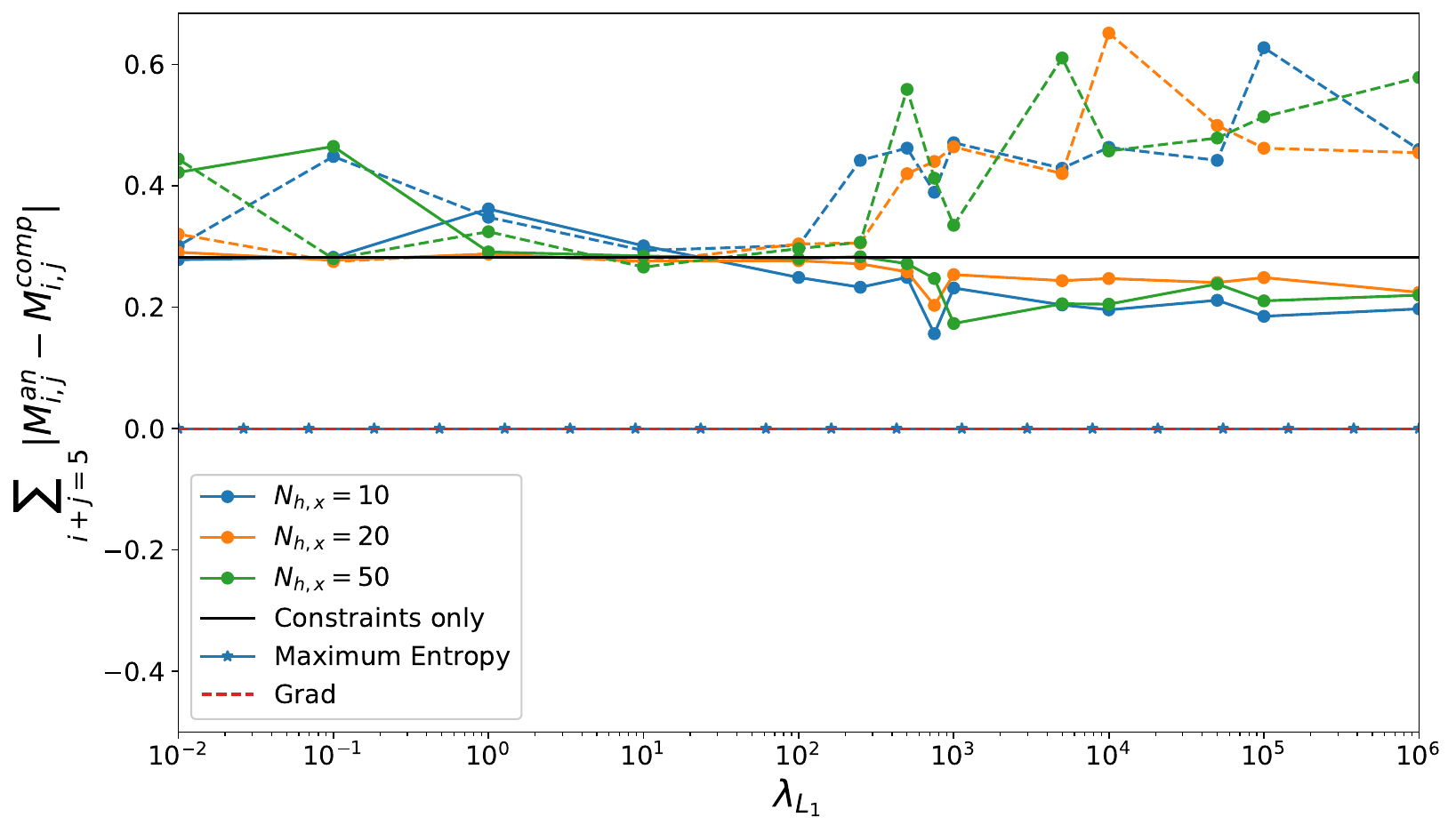}
  \caption{Number of distinct Diracs (left) and error in predicted moments of total order 5 (right) as a function of the sparsity-promoting regularization strength $\lambda_{L_1}$ for the case of $N=4$; Maxwell--Boltzmann distribution. Solid lines: $R_2$ regularization term, dashed lines: $R_1$ regularization term. Analytical value of $\sum_{i+j=N+1}|M_{i,j}^{an}|$ is 0.}
  \label{fig:mb4}
\end{figure}
From the right subplot of Figure~\ref{fig:mb3} we observe that using a low number of histogram bins leads to improved approximation quality compared to simply using a solution that satisfies the constraints. This is expected, as for higher numbers of histogram bins, it is less likely that any of the points end up in the same bin (since $N_p \ll N_{h,x} \times N_{h,y}$. As only the number of Diracs in a single bin (and not their velocities) affects the value of the approximation of the entropy, having coarser bins leads to the value of the entropy being more strongly influenced by the changes in the positions of the Diracs, as the number of them in bins is more likely to change. Naturally, use of too coarse bins also reduces the accuracy due to the coarse discretization of the integral.

Enforcing sparsity leads to higher errors in the predicted moments if the underlying distribution $f$ is not sparse. A minimum value of $\lambda_{L_1}$ of approximately 1000 is required to produce a sparse solution. As expected, optimizing directly for the Euclidean distance (which is used in the post-processing to find Diracs that are close) leads to a higher degree of sparsity than optimizing for the Manhattan distance. In addition, using the Manhattan distance in the sparsity term leads to larger errors (see Figure~\ref{fig:mb4}). For high values of  $\lambda_{L_1}$, the number of distinct Diracs is roughly equal to the number of moment constraints — 9 for $N=3$ and 14 for $N=4$. The number of moment constraints is lower by one than the total number of conserved moments, as the density constraint is enforced by construction of the ansatz~(\ref{eq:f-expansion}.

For the equilibrium case, both Grad's method and the classical maximum entropy approach recover the Maxwell--Boltzmann distribution exactly, and thus have 0 error.

\subsection{Druyvesteyn distribution}
Next, we consider the 2-dimensional Druyvesteyn distribution:
\begin{equation}
 f(v_1,v_2) = \frac{2}{\pi^2} \exp\left(-\frac{1}{\pi} \left(v_1^2 + v_2^2\right)^2 \right).
\end{equation}

\begin{figure}[h]
  \centering
  \includegraphics[width=.47\textwidth]{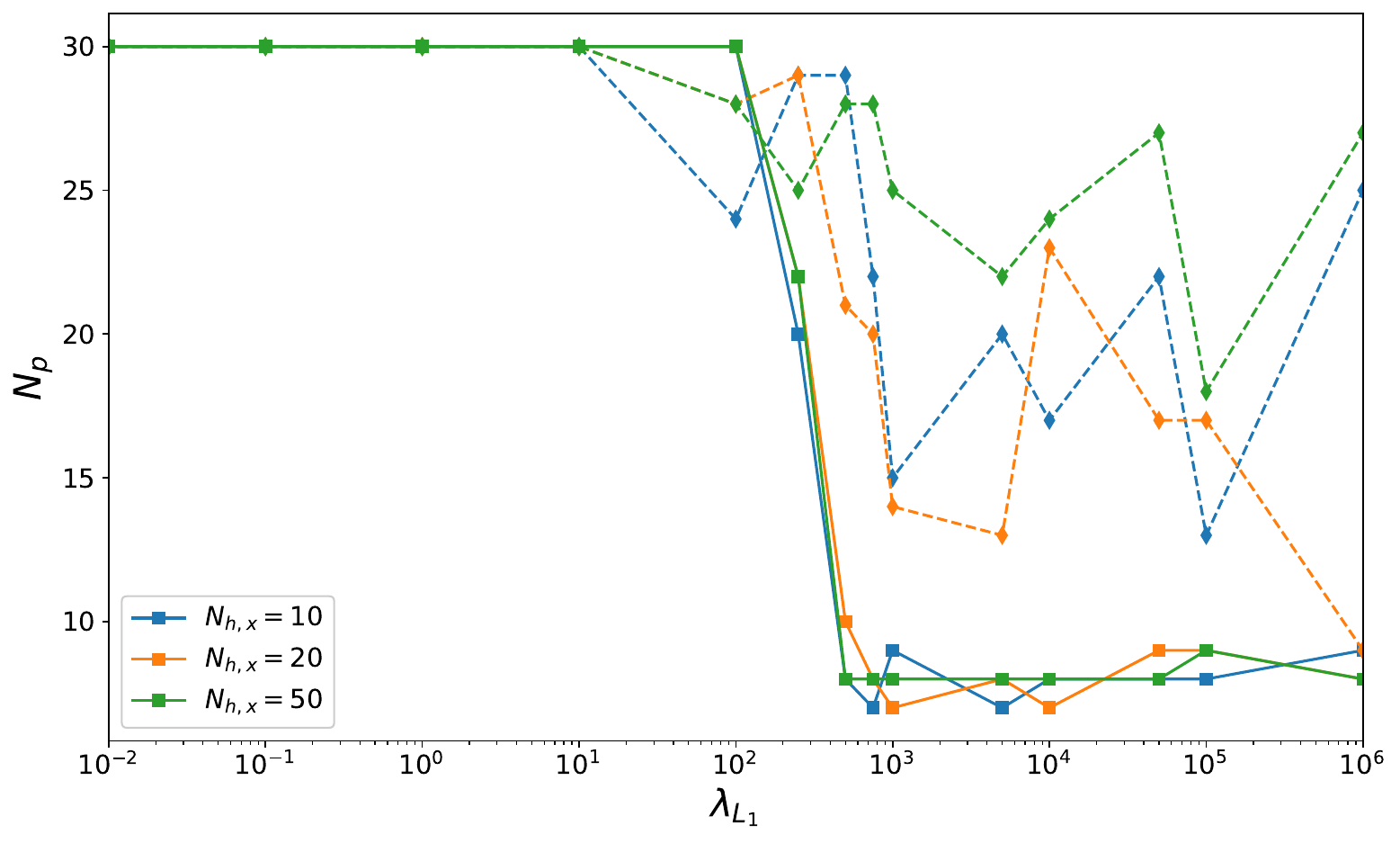}
  \includegraphics[width=.47\textwidth]{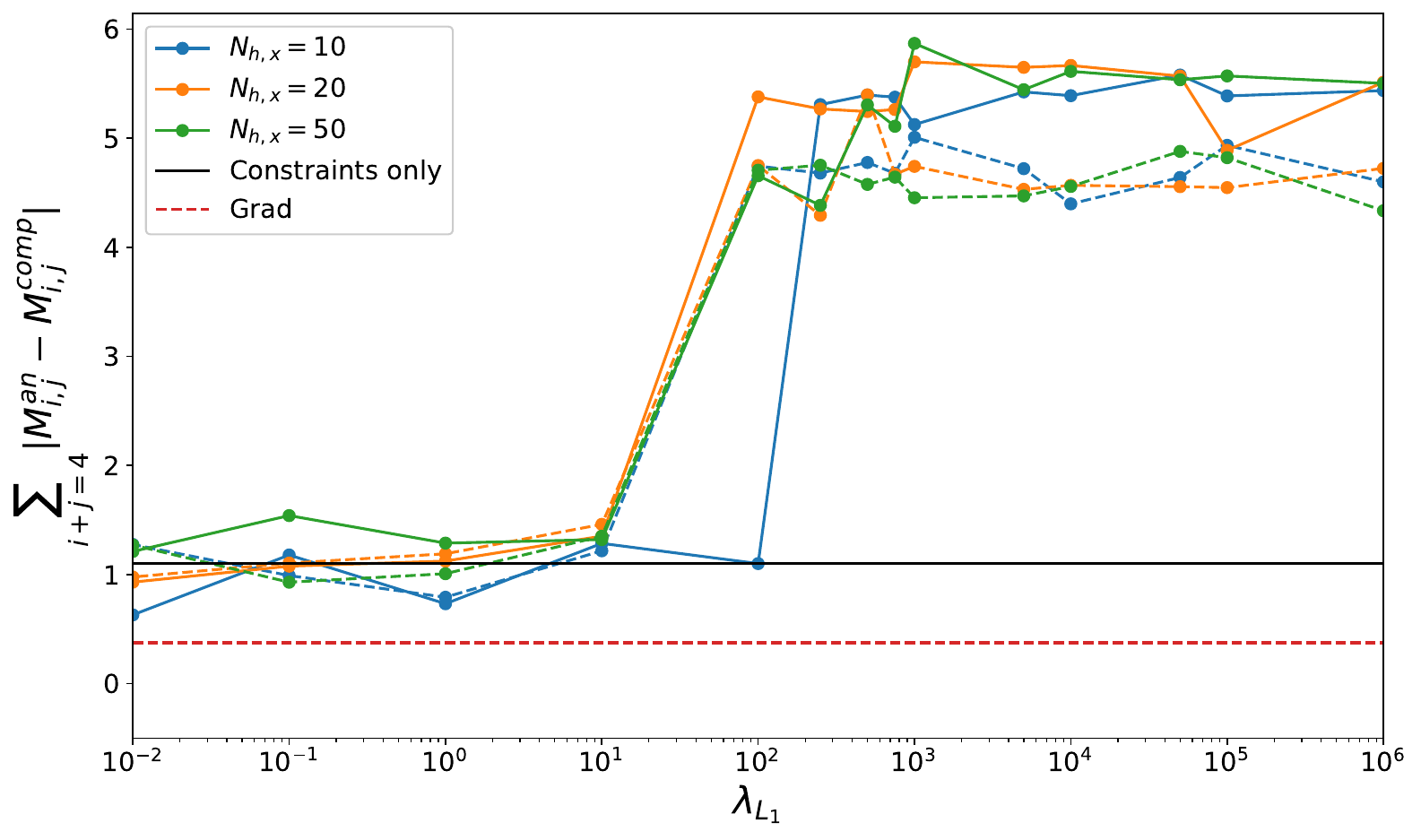}
  \caption{Number of distinct Diracs (left) and error in predicted moments of total order 4 (right) as a function of the sparsity-promoting regularization strength $\lambda_{L_1}$ for the case of $N=3$; Druyvesteyn distribution. Solid lines: $R_2$ regularization term, dashed lines: $R_1$ regularization term. Analytical value of $\sum_{i+j=N+1}|M_{i,j}^{an}|$ is approximately 1.3744.}
  \label{fig:dr3}
\end{figure}

Figure~\ref{fig:dr3} shows the number of distinct Diracs (left) and the error in the next predicted moments (right) for the case of $N=3$. For the odd-order case of the Druyvesteyn distribution, the Grad approach has lower error compared to the method developed in the present work; for the latter, as in the case of the Maxwell--Boltzmann distribution, placing more emphasis on sparsity (by taking larger values of $\lambda_1$) leads to higher errors.
Also similar to the previous case of the Maxwell--Boltzmann distribution, directly minimizing the sum of Euclidean distances between the Dirac deltas leads to a higher degree of sparsity than minimizing the sum of Manhattan distances. 

\begin{figure}[h]
  \centering
  \includegraphics[width=.47\textwidth]{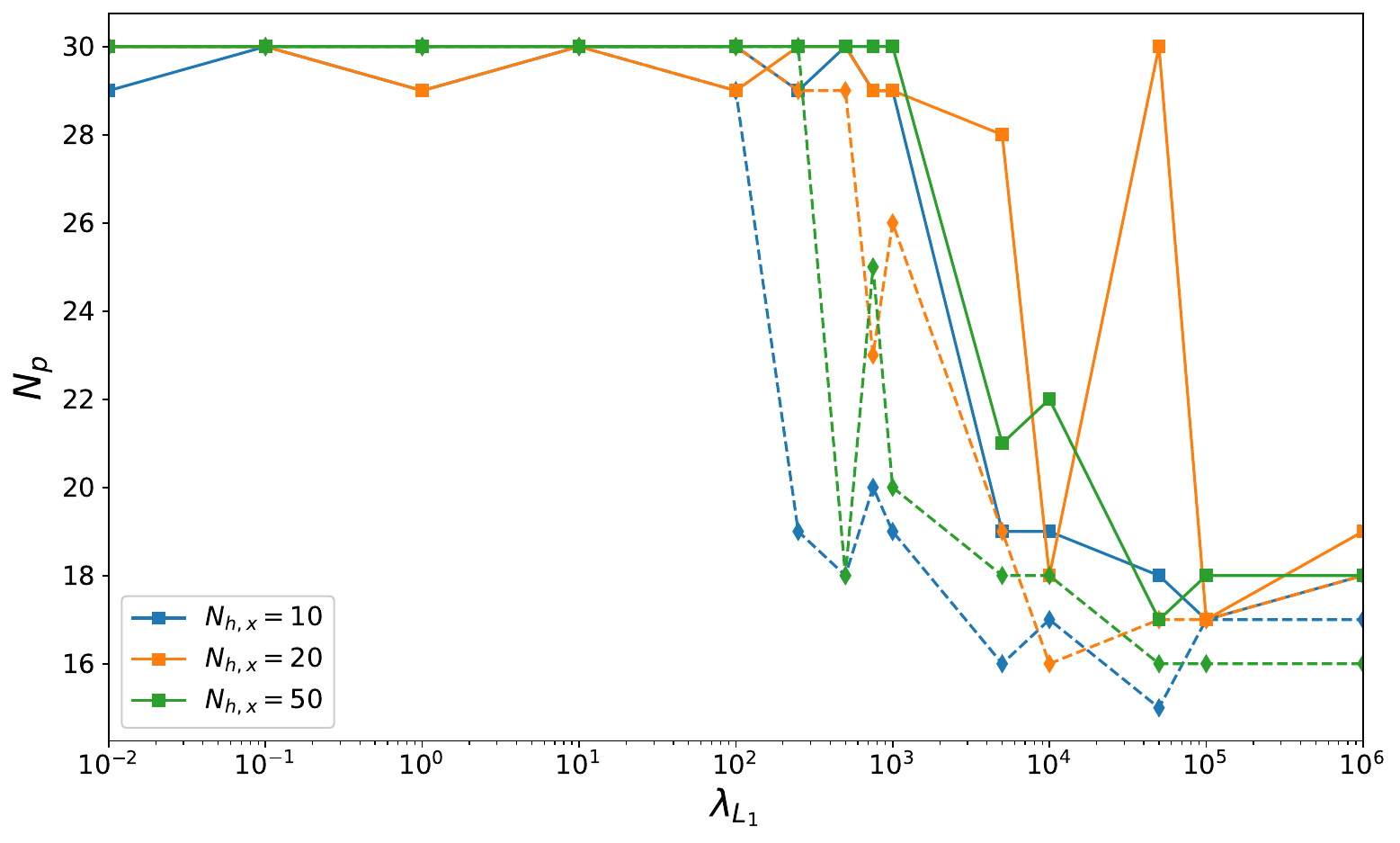}
  \includegraphics[width=.47\textwidth]{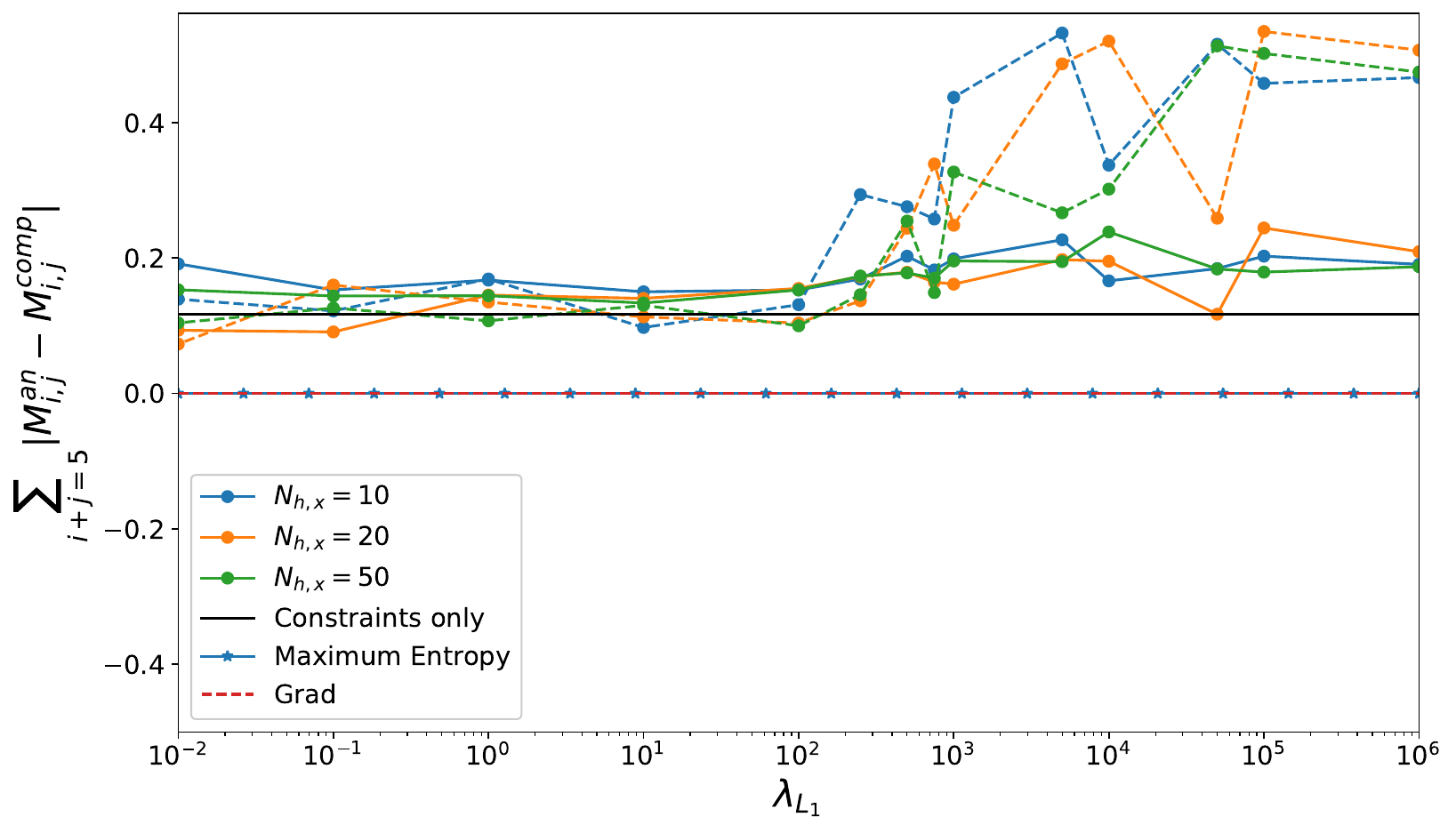}
  \caption{Number of distinct Diracs (left) and error in predicted moments of total order 5 (right) as a function of the sparsity-promoting regularization strength $\lambda_{L_1}$ for the case of $N=4$; Druyvesteyn distribution. Solid lines: $R_2$ regularization term, dashed lines: $R_1$ regularization term. Analytical value of $\sum_{i+j=N+1}|M_{i,j}^{an}|$ is 0.}
  \label{fig:dr4}
\end{figure}
For the case of $N=4$, use of the Grad and classical maximum entropy methods results in a symmetric distribution function, thus correctly predicting that the odd-order moments are 0, leading to them having no error in that case, as seen on Figure~\ref{fig:dr4}. For the approach developed in the present work, the use of the Manhattan distance in this case leads to a level of sparsity similar to that obtained via use of the Euclidean distance, but the error in the case of the Manhattan distance is noticeably larger.

\subsection{Bimodal distribution}\label{sec:bimodal}
Finally, we consider an asymmetric bimodal distribution that is a mixture of two Maxwell--Boltzmann distributions:
\begin{equation}
    f(v_1,v_2) = \frac{\rho_1}{\pi T_1} \exp\left(-\frac{(v_1-v_{x,1})^2 + (v_2-v_{y,2})^2}{T_1}\right) + \frac{\rho_2}{\pi T_2} \exp\left(-\frac{(v_1-v_{x,2})^2 + (v_2-v_{y,2})^2}{T_2}\right),
\end{equation}
with the following values of the parameters: $\rho_1=1/4$, $\rho_2=3/4$, $v_{x,1}=6/5$, $v_{y,1}=3/5$, $v_{x,2}=-2/5$, $v_{y,2}=-1/5$, $T_1=1/4$, $T_2=3/2$.

\begin{figure}[h]
  \centering
  \includegraphics[width=.47\textwidth]{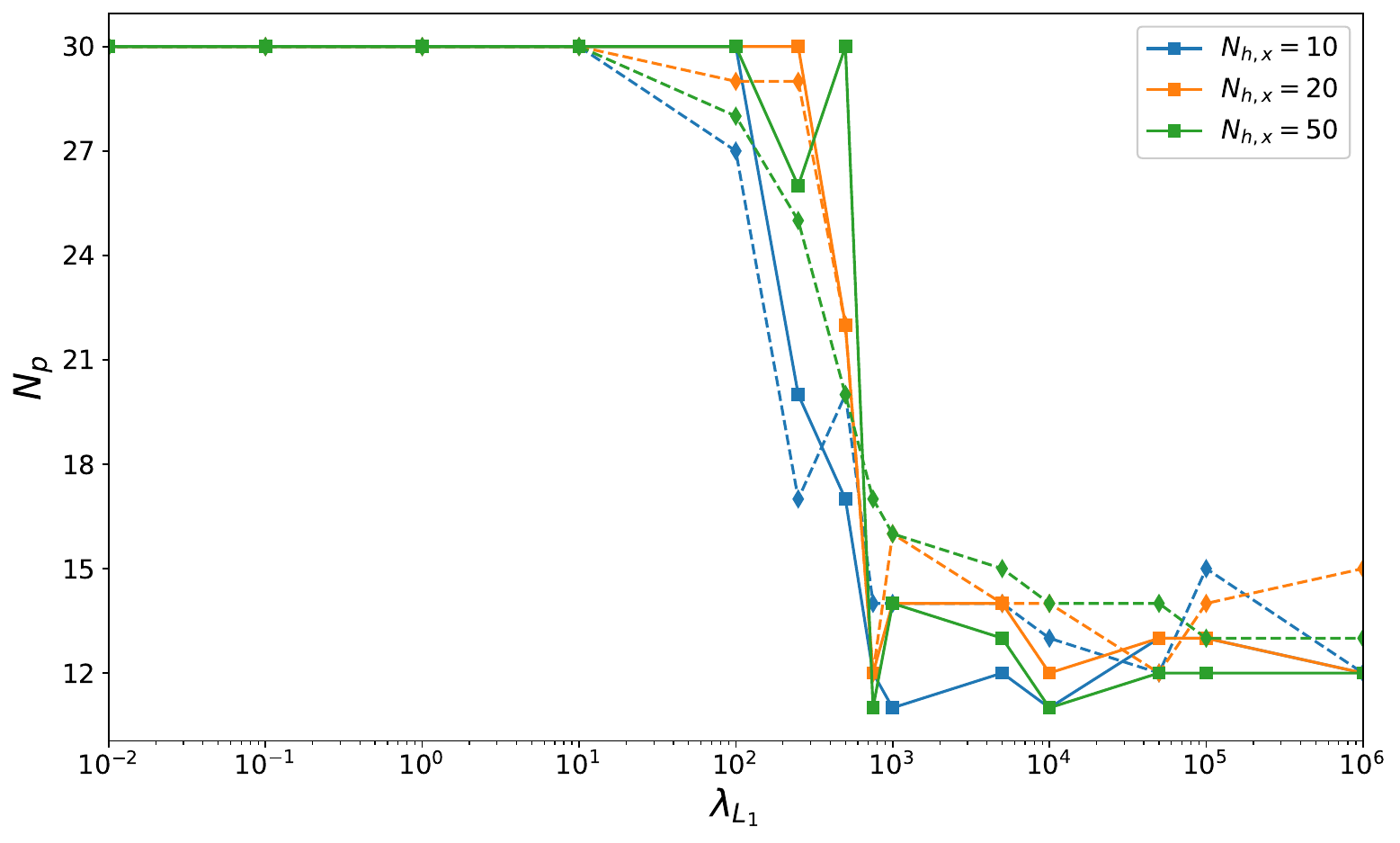}
  \includegraphics[width=.47\textwidth]{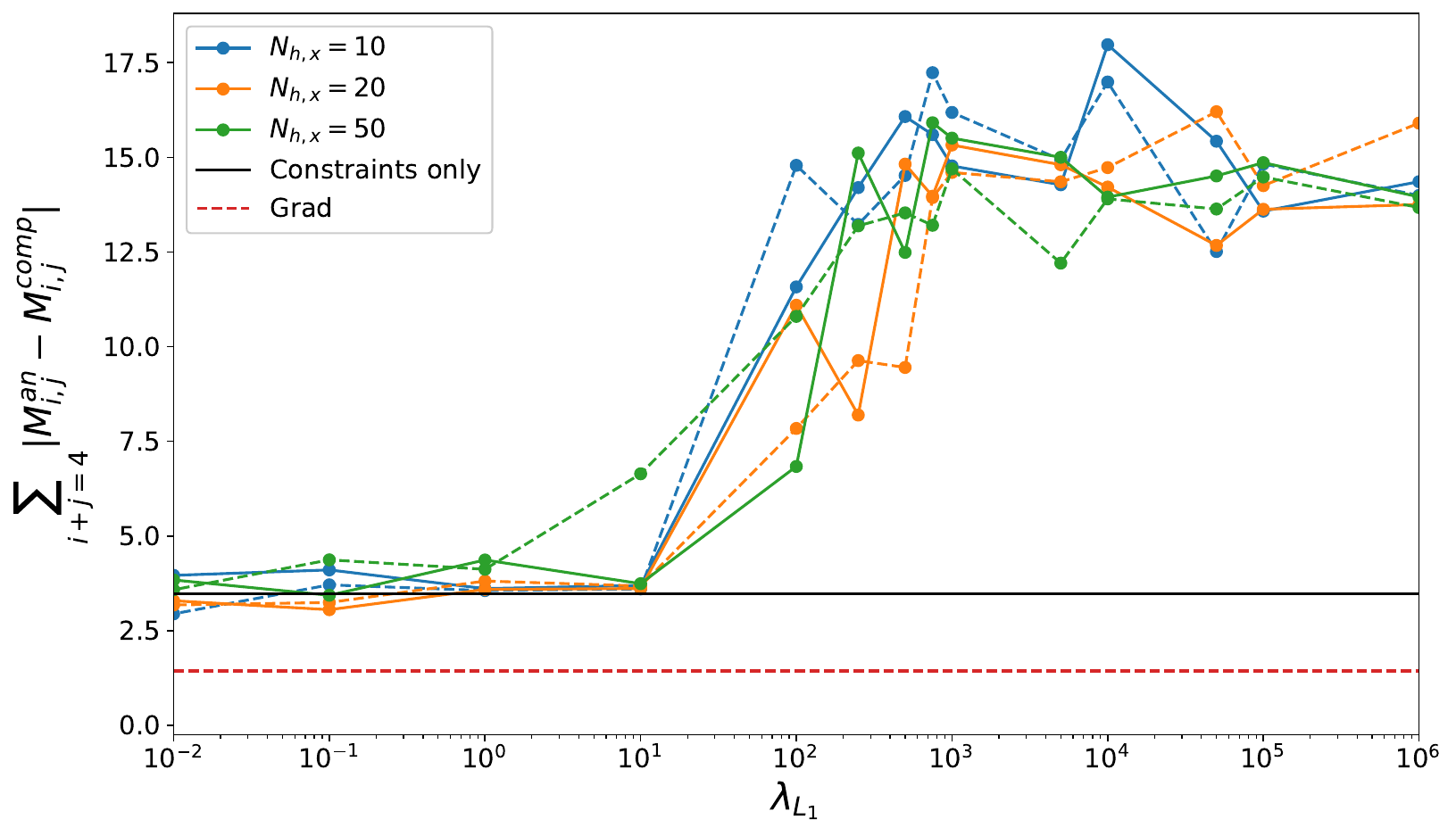}
  \caption{Number of distinct Diracs (left) and error in predicted moments of total order 4 (right) as a function of the sparsity-promoting regularization strength $\lambda_{L_1}$ for the case of $N=3$; bimodal distribution. Solid lines: $R_2$ regularization term, dashed lines: $R_1$ regularization term. Analytical value of $\sum_{i+j=N+1}|M_{i,j}^{an}|$ is approximately 5.6083.}
  \label{fig:bim3}
\end{figure}

Figure~\ref{fig:bim3} shows the sparsity of the solution (left) and the error in the next predicted moments (right) for the case of $N=3$. The behaviour of the solution with regards to the choice of $\lambda_1$ and the distance metric used is similar to that of the previous cases. The Grad closure performs better than the closure developed in the present work for this case, and the impact of minimizing entropy as compared to simply solving for the moment constraints is negligible.

\begin{figure}[h]
  \centering
  \includegraphics[width=.47\textwidth]{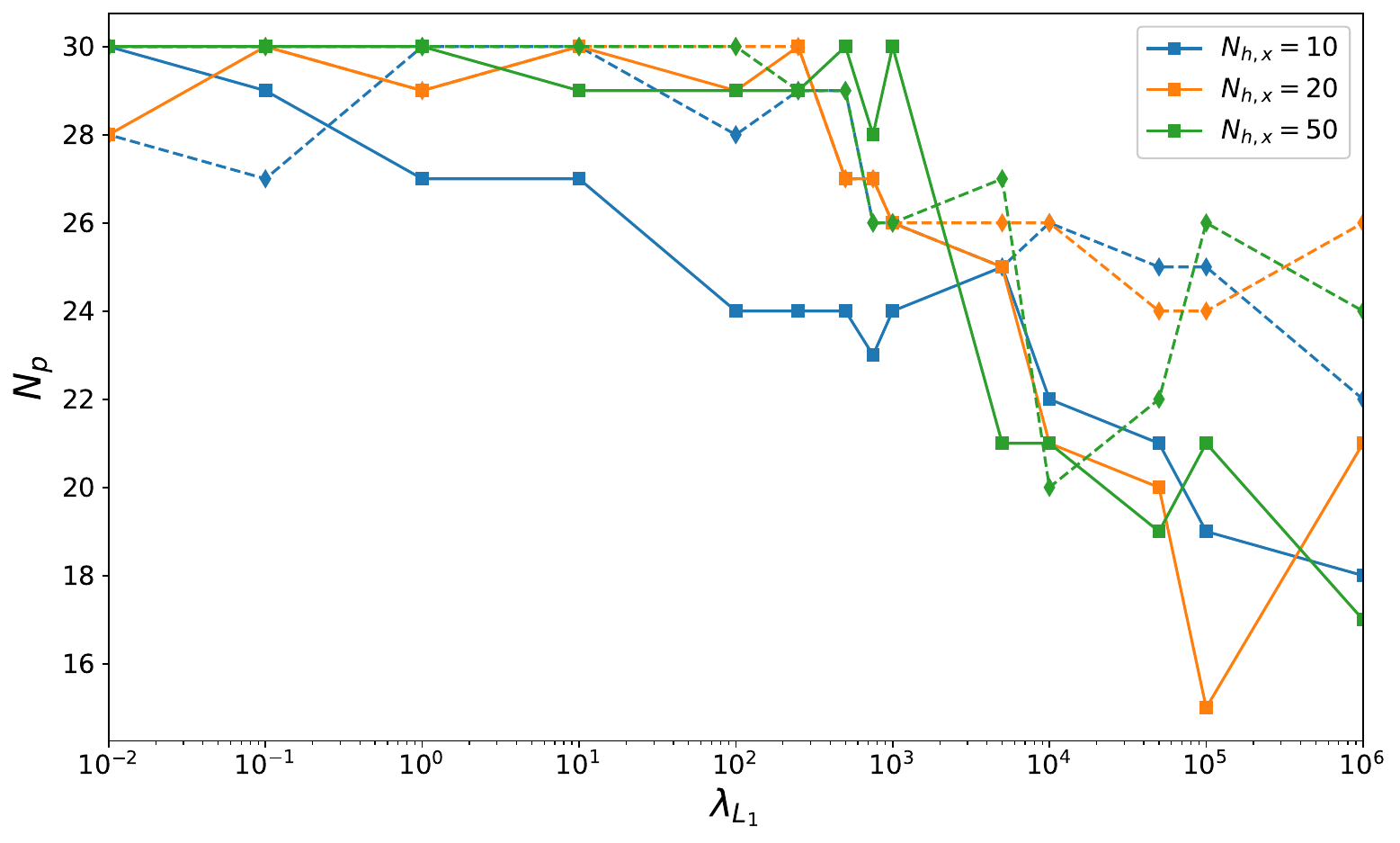}
  \includegraphics[width=.47\textwidth]{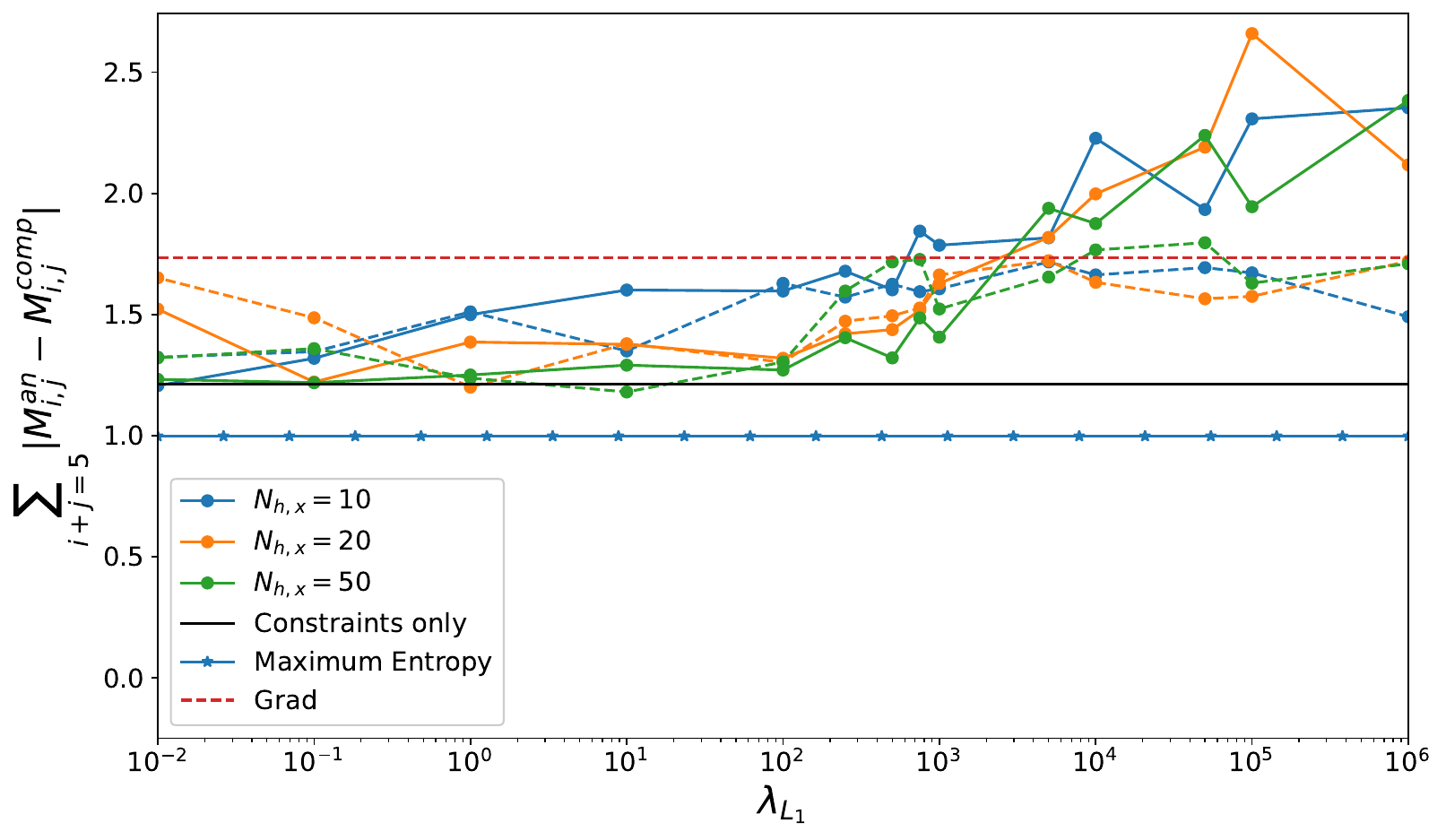}
  \caption{Number of distinct Diracs (left) and error in predicted moments of total order 5 (right) as a function of the sparsity-promoting regularization strength $\lambda_{L_1}$ for the case of $N=4$; bimodal distribution. Solid lines: $R_2$ regularization term, dashed lines: $R_1$ regularization term. Analytical value of $\sum_{i+j=N+1}|M_{i,j}^{an}|$ is approximately 3.8443.}
  \label{fig:bim4}
\end{figure}

For the case of $N=4$, as shown on Fig.~\ref{fig:bim4}, the Grad closure performs worse than the closure developed in the present work, unless a very sparse solution is produced by the latter. The classical maximum entropy approach provides the best results. Again, for this case, in the newly developed approach, the impact of minimizing the histogram-approximated entropy as compared to simply solving for the moment constraints is negligible; however, the overall error is comparable to that of the classical maximum entropy approach in the limit of small $\lambda_1$.

 \section{Conclusions}
A new  approach has been developed for the moment closure problem, which is based on a sparse reconstruction of the underlying distribution obtained via an optimization of an approximated entropy. The method has been applied to several model distributions arising
in kinetic theory and compared to the classical closure method of Grad and the classical maximum entropy approach. The impact of the choice of the distance metric used in the regularization term has been studied numerically, along with the role of the regularization term itself. The approach has been shown to have comparable error to existing methods in the limit of no regularization, and exhibits the expected accuracy-sparsity trade-off for smooth distribution functions. Moreover, it guarantees non-negativity of the distribution function by construction, and is robust in the limit of the realizability boundary.

Future work will focus on the improvement of the numerical stability and performance of the method, its application of the method to transport problems, as well as consider other ansatz possibilities (such as~\cite{yuan2012extended}) and the use of data-driven approaches for faster solution of the optimization problem (similar to~\cite{sadr2020gaussian}).

\section*{Acknowledgments}
The authors thank the Deutsche Forschungsgemeinschaft (DFG, German Research Foundation) for the financial support through 442047500/SFB1481 within the projects B04 (Sparsity fördernde Muster in kinetischen Hierarchien), B05 (Sparsifizierung zeitabhängiger Netzwerkflußprobleme mittels diskreter Optimierung) and B06 (Kinetische Theorie trifft algebraische Systemtheorie).


\begin{thebibliography}{53}
\providecommand{\natexlab}[1]{#1}
\providecommand{\url}[1]{\texttt{#1}}
\expandafter\ifx\csname urlstyle\endcsname\relax
  \providecommand{\doi}[1]{doi: #1}\else
  \providecommand{\doi}{doi: \begingroup \urlstyle{rm}\Url}\fi

\bibitem[Abdelmalik and Van~Brummelen(2016)]{abdelmalik2016moment}
M.~Abdelmalik and E.~Van~Brummelen.
\newblock Moment closure approximations of the {B}oltzmann equation based on
  $\varphi$-divergences.
\newblock \emph{J. Stat. Phys.}, 164\penalty0 (1):\penalty0 77--104, 2016.

\bibitem[Aheizer and Krein(1962)]{Aheizer1962}
N.~Aheizer and M.~Krein.
\newblock \emph{Some Questions in the Theory of Moments}, volume~2 of
  \emph{Translations of Mathematical Monographs}.
\newblock American Mathematical Society, 1962.

\bibitem[Alldredge et~al.(2019)Alldredge, Frank, and
  Hauck]{alldredge2019regularized}
G.~W. Alldredge, M.~Frank, and C.~D. Hauck.
\newblock A regularized entropy-based moment method for kinetic equations.
\newblock \emph{SIAM J. Appl. Math.}, 79\penalty0 (5):\penalty0 1627--1653,
  2019.

\bibitem[B{\"o}hmer and Torrilhon(2020)]{bohmer2020entropic}
N.~B{\"o}hmer and M.~Torrilhon.
\newblock Entropic quadrature for moment approximations of the {Boltzmann-BGK}
  equation.
\newblock \emph{J. Comput. Phys.}, 401:\penalty0 108992, 2020.

\bibitem[Cai(2021)]{cai2021moment}
Z.~Cai.
\newblock Moment method as a numerical solver: challenge from shock structure
  problems.
\newblock \emph{J. Comput. Phys.}, 444:\penalty0 110593, 2021.

\bibitem[Cai and Torrilhon(2019)]{cai2019holway}
Z.~Cai and M.~Torrilhon.
\newblock On the {Holway-Weiss} debate: Convergence of the
  {Grad}-moment-expansion in kinetic gas theory.
\newblock \emph{Phys. Fluids}, 31\penalty0 (12), 2019.

\bibitem[Cercignani(1988)]{cercignani1988}
C.~Cercignani.
\newblock \emph{The {B}oltzmann Equation and Its Applications}.
\newblock Springer New York, NY, 1988.

\bibitem[Cercignani(2005)]{struchtrup2005}
C.~Cercignani.
\newblock \emph{Macroscopic Transport Equations for Rarefied Gas Flows}.
\newblock Springer Berlin, Heidelberg, 2005.

\bibitem[Chalons et~al.(2010)Chalons, Kah, and Massot]{chalons2010beyond}
C.~Chalons, D.~Kah, and M.~Massot.
\newblock Beyond pressureless gas dynamics: quadrature-based velocity moment
  models.
\newblock \emph{arXiv preprint arXiv:1011.2974}, 2010.

\bibitem[Crestetto et~al.(2012)Crestetto, Crouseilles, and
  Lemou]{crestetto2012kinetic}
A.~Crestetto, N.~Crouseilles, and M.~Lemou.
\newblock Kinetic/fluid micro-macro numerical schemes for
  {V}lasov-{P}oisson-{BGK} equation using particles.
\newblock \emph{Kinet. Relat. Models}, 5\penalty0 (4):\penalty0 787--816, 2012.

\bibitem[Desjardins et~al.(2008)Desjardins, Fox, and
  Villedieu]{desjardins2008quadrature}
O.~Desjardins, R.~O. Fox, and P.~Villedieu.
\newblock A quadrature-based moment method for dilute fluid-particle flows.
\newblock \emph{J. Comput. Phys.}, 227\penalty0 (4):\penalty0 2514--2539, 2008.

\bibitem[Dimarco and Pareschi(2014)]{dimarco2014numerical}
G.~Dimarco and L.~Pareschi.
\newblock Numerical methods for kinetic equations.
\newblock \emph{Acta Numer.}, 23:\penalty0 369--520, 2014.

\bibitem[Eftimie(2018)]{Eftimie2018}
R.~Eftimie.
\newblock \emph{Multi-Dimensional Transport Equations}, pages 153--193.
\newblock Springer International Publishing, Cham, 2018.

\bibitem[Foucart and Koslicki(2014)]{foucart2014sparse}
S.~Foucart and D.~Koslicki.
\newblock Sparse recovery by means of nonnegative least squares.
\newblock \emph{IEEE Signal Process. Lett.}, 21\penalty0 (4):\penalty0
  498--502, 2014.

\bibitem[Fox(2008)]{fox2008quadrature}
R.~O. Fox.
\newblock A quadrature-based third-order moment method for dilute gas-particle
  flows.
\newblock \emph{J. Comput. Phys.}, 227\penalty0 (12):\penalty0 6313--6350,
  2008.

\bibitem[Fox and Laurent(2022)]{FoxLaurent}
R.~O. Fox and F.~Laurent.
\newblock Hyperbolic quadrature method of moments for the one-dimensional
  kinetic equation.
\newblock \emph{SIAM J. Appl. Math.}, 82\penalty0 (2):\penalty0 750--771, 2022.

\bibitem[Fox et~al.(2018)Fox, Laurent, and Vi{\'e}]{fox2018conditional}
R.~O. Fox, F.~Laurent, and A.~Vi{\'e}.
\newblock Conditional hyperbolic quadrature method of moments for kinetic
  equations.
\newblock \emph{J. Comput. Phys.}, 365:\penalty0 269--293, 2018.

\bibitem[Fox et~al.(2023)Fox, Laurent, and Passalacqua]{fox2023generalized}
R.~O. Fox, F.~Laurent, and A.~Passalacqua.
\newblock The generalized quadrature method of moments.
\newblock \emph{J. Aerosol Sci.}, 167:\penalty0 106096, 2023.

\bibitem[Gillespie(2009)]{gillespie2009moment}
C.~S. Gillespie.
\newblock Moment-closure approximations for mass-action models.
\newblock \emph{IET Syst. Biol.}, 3\penalty0 (1):\penalty0 52--58, 2009.

\bibitem[Gonoskov(2022)]{gonoskov2022agnostic}
A.~Gonoskov.
\newblock Agnostic conservative down-sampling for optimizing statistical
  representations and {PIC} simulations.
\newblock \emph{Comput. Phys. Commun.}, 271:\penalty0 108200, 2022.

\bibitem[Grad(1949)]{grad2kinetic}
H.~Grad.
\newblock On the kinetic theory of rarefied gases.
\newblock \emph{Commun. Pure Appl. Math}, 2\penalty0 (331), 1949.

\bibitem[Hamburger(1944)]{hamburger1944hermitian}
H.~L. Hamburger.
\newblock Hermitian transformations of deficiency-index (1, 1), {J}acobi
  matrices and undetermined moment problems.
\newblock \emph{Am. J. Math.}, 66\penalty0 (4):\penalty0 489--522, 1944.

\bibitem[Herty et~al.(2020)Herty, Puppo, Roncoroni, and Visconti]{herty2020bgk}
M.~Herty, G.~Puppo, S.~Roncoroni, and G.~Visconti.
\newblock The {BGK} approximation of kinetic models for traffic.
\newblock \emph{Kinet. Relat. Models}, 13\penalty0 (2):\penalty0 279--307,
  2020.

\bibitem[Huang et~al.(2020)Huang, Li, and Yong]{huang2020stability}
Q.~Huang, S.~Li, and W.-A. Yong.
\newblock Stability analysis of quadrature-based moment methods for kinetic
  equations.
\newblock \emph{SIAM J. Appl. Math.}, 80\penalty0 (1):\penalty0 206--231, 2020.

\bibitem[Koellermeier and Rominger(2020)]{koellermeier2020analysis}
J.~Koellermeier and M.~Rominger.
\newblock Analysis and numerical simulation of hyperbolic shallow water moment
  equations.
\newblock \emph{Commun. Comput. Phys.}, 28\penalty0 (3):\penalty0 1038--1084,
  2020.

\bibitem[Koellermeier et~al.(2014)Koellermeier, Schaerer, and
  Torrilhon]{koellermeier2014framework}
J.~Koellermeier, R.~P. Schaerer, and M.~Torrilhon.
\newblock A framework for hyperbolic approximation of kinetic equations using
  quadrature-based projection methods.
\newblock \emph{Kinet. Relat. Models}, 7\penalty0 (3), 2014.

\bibitem[Levermore(1996)]{levermore1996moment}
C.~D. Levermore.
\newblock Moment closure hierarchies for kinetic theories.
\newblock \emph{J. Stat. Phys.}, 83:\penalty0 1021--1065, 1996.

\bibitem[Levermore(1997)]{levermore1997entropy}
C.~D. Levermore.
\newblock Entropy-based moment closures for kinetic equations.
\newblock \emph{Transp. Theor. Stat.}, 26\penalty0 (4-5):\penalty0 591--606,
  1997.

\bibitem[Lubin et~al.(2023)Lubin, Dowson, {Dias Garcia}, Huchette, Legat, and
  Vielma]{Lubin2023}
M.~Lubin, O.~Dowson, J.~{Dias Garcia}, J.~Huchette, B.~Legat, and J.~P. Vielma.
\newblock {JuMP} 1.0: {R}ecent improvements to a modeling language for
  mathematical optimization.
\newblock \emph{Math. Program. Comput.}, 2023.

\bibitem[Marques~Jr and M{\'e}ndez(2013)]{marques2013kinetic}
W.~Marques~Jr and A.~M{\'e}ndez.
\newblock On the kinetic theory of vehicular traffic flow: {C}hapman--{E}nskog
  expansion versus {G}rad’s moment method.
\newblock \emph{Phys. A: Stat. Mech. Appl.}, 392\penalty0 (16):\penalty0
  3430--3440, 2013.

\bibitem[Martin and Cambier(2016)]{martin2016octree}
R.~S. Martin and J.-L. Cambier.
\newblock Octree particle management for {DSMC} and {PIC} simulations.
\newblock \emph{J. Comput. Phys.}, 327:\penalty0 943--966, 2016.

\bibitem[McDonald and Torrilhon(2013)]{mcdonald2013affordable}
J.~McDonald and M.~Torrilhon.
\newblock Affordable robust moment closures for {CFD} based on the
  maximum-entropy hierarchy.
\newblock \emph{J. Comput. Phys.}, 251:\penalty0 500--523, 2013.

\bibitem[McGraw(1997)]{mcgraw1997description}
R.~McGraw.
\newblock Description of aerosol dynamics by the quadrature method of moments.
\newblock \emph{Aerosol Sci. Tech.}, 27\penalty0 (2):\penalty0 255--265, 1997.

\bibitem[Milbrandt and Yau(2005)]{milbrandt2005multimoment}
J.~Milbrandt and M.~Yau.
\newblock A multimoment bulk microphysics parameterization. part {II}: A
  proposed three-moment closure and scheme description.
\newblock \emph{J. Atmos. Sci.}, 62\penalty0 (9):\penalty0 3065--3081, 2005.

\bibitem[Modest and Mazumder(2021)]{modest2021radiative}
M.~F. Modest and S.~Mazumder.
\newblock \emph{Radiative heat transfer}.
\newblock Academic press, 2021.

\bibitem[Oblapenko(2024)]{oblapenko2024non}
G.~Oblapenko.
\newblock A non-negative least squares-based approach for moment-preserving
  particle merging.
\newblock \emph{arXiv preprint arXiv:2412.12354}, 2024.

\bibitem[Oblapenko et~al.(2025)Oblapenko, Torrilhon, and
  Herty]{oblapenko2025reprorepo}
G.~Oblapenko, M.~Torrilhon, and M.~Herty.
\newblock Reproducibility repository for "{S}parse reconstruction of
  multi-dimensional kinetic distributions".
\newblock \url{https://github.com/knstmrd/paper-sparse_reconstruction_kdf}, 3
  2025.

\bibitem[Powell(1994)]{powell1994direct}
M.~J. Powell.
\newblock \emph{A direct search optimization method that models the objective
  and constraint functions by linear interpolation}.
\newblock Springer, 1994.

\bibitem[Sadr et~al.(2020)Sadr, Torrilhon, and Gorji]{sadr2020gaussian}
M.~Sadr, M.~Torrilhon, and M.~H. Gorji.
\newblock Gaussian process regression for maximum entropy distribution.
\newblock \emph{J. Comput. Phys.}, 418:\penalty0 109644, 2020.

\bibitem[Sadr et~al.(2024)Sadr, Hadjiconstantinou, and
  Gorji]{sadr2024wasserstein}
M.~Sadr, N.~G. Hadjiconstantinou, and M.~H. Gorji.
\newblock Wasserstein-penalized entropy closure: A use case for stochastic
  particle methods.
\newblock \emph{J. Comput. Phys.}, 511:\penalty0 113066, 2024.

\bibitem[Schm{\"u}dgen(2017)]{schmuedgen2017moment}
K.~Schm{\"u}dgen.
\newblock \emph{The Moment Problem}.
\newblock Graduate Texts in Mathematics. Springer International Publishing,
  2017.

\bibitem[Shohat and Tamarkin(1945)]{Shohat1945problem}
J.~Shohat and J.~Tamarkin.
\newblock \emph{The problem of moments}.
\newblock AMS, Providence, 1945.

\bibitem[Singh and Hespanha(2006)]{singh2006moment}
A.~Singh and J.~P. Hespanha.
\newblock Moment closure techniques for stochastic models in population
  biology.
\newblock In \emph{2006 American Control Conference}. IEEE, 2006.

\bibitem[Slawski and Hein(2013)]{slawski2013non}
M.~Slawski and M.~Hein.
\newblock Non-negative least squares for high-dimensional linear models:
  Consistency and sparse recovery without regularization.
\newblock \emph{Electron. J. Statist.}, 7:\penalty0 3004, 2013.

\bibitem[Struchtrup(1998)]{struchtrup1998number}
H.~Struchtrup.
\newblock On the number of moments in radiative transfer problems.
\newblock \emph{Ann. Phys. (N. Y.)}, 266\penalty0 (1):\penalty0 1--26, 1998.

\bibitem[Struchtrup and Torrilhon(2003)]{struchtrup2003regularization}
H.~Struchtrup and M.~Torrilhon.
\newblock Regularization of {G}rad’s 13 moment equations: Derivation and
  linear analysis.
\newblock \emph{Phys. Fluids}, 15\penalty0 (9):\penalty0 2668--2680, 2003.

\bibitem[Torrilhon(2016)]{torrilhon2016modeling}
M.~Torrilhon.
\newblock Modeling nonequilibrium gas flow based on moment equations.
\newblock \emph{Annu. Rev. Fluid Mech.}, 48:\penalty0 429--458, 2016.

\bibitem[Van~Cappellen et~al.(2021)Van~Cappellen, Vetrano, and
  Laboureur]{van2021higher}
M.~Van~Cappellen, M.~R. Vetrano, and D.~Laboureur.
\newblock Higher order hyperbolic quadrature method of moments for solving
  kinetic equations.
\newblock \emph{J. Comput. Phys.}, 436:\penalty0 110280, 2021.

\bibitem[Wheeler(1974)]{wheeler1974modified}
J.~C. Wheeler.
\newblock Modified moments and {G}aussian quadratures.
\newblock \emph{The Rocky Mountain Journal of Mathematics}, 4\penalty0
  (2):\penalty0 287--296, 1974.

\bibitem[Yilmaz et~al.(2024)Yilmaz, Oblapenko, and
  Torrilhon]{yilmaz2024nonlinear}
E.~Yilmaz, G.~Oblapenko, and M.~Torrilhon.
\newblock On nonlinear closures for moment equations based on orthogonal
  polynomials.
\newblock \emph{arXiv preprint arXiv:2407.05894}, 2024.

\bibitem[Yuan and Fox(2011)]{yuan2011conditional}
C.~Yuan and R.~O. Fox.
\newblock Conditional quadrature method of moments for kinetic equations.
\newblock \emph{J. Comput. Phys.}, 230\penalty0 (22):\penalty0 8216--8246,
  2011.

\bibitem[Yuan et~al.(2012)Yuan, Laurent, and Fox]{yuan2012extended}
C.~Yuan, F.~Laurent, and R.~Fox.
\newblock An extended quadrature method of moments for population balance
  equations.
\newblock \emph{J. Aerosol Sci.}, 51:\penalty0 1--23, 2012.

\bibitem[Zhang(2023)]{Zhang_2023}
Z.~Zhang.
\newblock {PRIMA: Reference Implementation for Powell's Methods with
  Modernization and Amelioration}.
\newblock https://github.com/libprima/prima, 2023.

\end{thebibliography}
\end{document}